\documentclass[letterpaper,twocolumn,10pt]{article}
\usepackage{usenix2019_v3}

\usepackage{tikz}
\usepackage{amsmath}
\usepackage{xspace}
\usepackage{amsfonts}
\usepackage{subcaption}
\usepackage{multirow}
\usepackage{algorithm}
\usepackage[noend]{algpseudocode}
\usepackage{enumitem}
\usepackage{titlesec}
\usepackage[T1]{fontenc}
\usepackage{inconsolata}
\usepackage{makecell}
\usepackage{comment}
\usepackage{makecell}
\usepackage[most]{tcolorbox}

\usepackage{amsthm}

\newtheoremstyle{tight}  
  {2pt}                  
  {2pt}                  
  {\itshape}             
  {}                     
  {\bfseries}            
  {}                    
  { }                    
  {}                     

\theoremstyle{tight}
\newtheorem{theorem}{Theorem}
\newtheorem{lemma}{Lemma}

\newcommand{\PHB}[1]{\noindent\textbf{#1}}
\newcommand{\PHM}[1]{\vspace{.4em}\noindent\textbf{#1}} 
\newcommand{\SystemName}{\textsc{Adaptra}\xspace}
\newcommand{\GreyHound}{Falcon}

\begin{document}
\setlength{\floatsep}{3pt plus 2pt minus 2pt}
\setlength{\textfloatsep}{3pt plus 2pt minus 2pt}
\setlength{\intextsep}{3pt plus 2pt minus 2pt}
\setlength{\abovecaptionskip}{3pt plus 1pt minus 1pt}
\setlength{\belowcaptionskip}{3pt plus 1pt minus 1pt}
\setlength{\abovedisplayskip}{3pt}
\setlength{\belowdisplayskip}{3pt}


\date{}


\title{ \Large \bf \SystemName: Straggler-Resilient Hybrid-Parallel Training with Pipeline Adaptation}

\author{
  Tianyuan Wu$^{\dagger}$\thanks{Equal contribution.}, 
  Lunxi Cao$^{\dagger}$\footnotemark[1],
  Hanfeng Lu$^{\dagger}$,
  Xiaoxiao Jiang$^{\dagger}$,
  Yinghao Yu$^{\S}$,
  Siran Yang$^{\S}$,\\
  Guodong Yang$^{\S}$, 
  Jiamang Wang$^{\S}$, 
  Lin Qu$^{\S}$, 
  Liping Zhang$^{\S}$,
  Wei Wang$^{\dagger}$ \\
  $^{\dagger}$Hong Kong University of Science and Technology \\
  $^{\S}$Alibaba Group
}

\maketitle


\begin{abstract}

Training large Deep Neural Network (DNN) models at scale often encounters
straggler issues, mostly in communications due to network congestion,
RNIC/switch defects, or topological asymmetry. Under advanced pipeline
parallelism, even minor communication delays can induce significant training
slowdowns. This occurs because (1) slow communication disrupts the pipeline
schedule, creating cascading ``bubbles'' in a domino effect, and (2) current
GPU kernel scheduling is susceptible to head-of-line blocking, where slow
communication blocks subsequent computations, further adding to these
bubbles. To address these challenges, we present \SystemName{}, a
straggler-resilient training system with two key optimizations. First, it
optimally adapts the pipeline schedule in the presence of stragglers to
absorb communication delays without inducing cascading bubbles, using a
simple yet effective algorithm guided by an analytical model. Second, upon
detecting slow communication, \SystemName{} offloads communication operations
from GPU to host memory and utilizes CPU-side RDMA for data transfer. This
eliminates head-of-line blocking as subsequent computation kernels can be
scheduled immediately on GPUs. Together, these optimizations effectively
reduce pipeline stalls in the presence of communication stragglers,
improving the training iteration time by 1.2-3.5$\times$ in our
experiments under various settings.


\end{abstract}


\section{Introduction}
\label{sec:intro}

The rise of large Deep Neural Network (DNN) models has ushered in the golden
age of Artificial Intelligence, leading to breakthroughs in
applications that would have been considered science fiction even a few years
ago~\cite{villalobos2022machine, zhang2022opt, sora, achiam2023gpt,
team2023gemini, guo2025deepseek}. Training large DNN models typically
requires combining tensor, data, and pipeline parallelism strategies across
thousands of GPUs~\cite{narayanan2021megatron, jiang2024megascale,
shoeybi2019megatron, zheng2022alpa}, where providing high-throughput,
low-latency communication is critical to enhancing the training performance~\cite
{dong2021accl, cao2024crux,liu2024deepseek}.


However, at this scale, communication stragglers, manifested as slow links
with extended pairwise transmission delays, are frequent~\cite
{dubey2024llama, qian2024alibaba} and can have a significant performance
impact~\cite{cui2025xputimer, rajasekaran2024cassini,
wu2024falcon}. Particularly in multi-tenant environments, jobs occupying a
large number of GPUs frequently suffer from communication-induced slowdowns
during life cycles~\cite{cao2024crux, wu2024falcon}. These stragglers
originate predominantly from transient network congestion~\cite
{cao2024crux, wu2024falcon} but can also persist due to hardware defects
(e.g., RNIC or switch failures) or network topological asymmetry~\cite
{xiong2024superbench}. The presence of communication stragglers, even on a
single link, slows down the entire training job due to frequent
synchronizations necessitated by hybrid-parallelism strategies, causing up to 90\%
throughput degradation in production clusters~\cite
{cui2025xputimer, rajasekaran2024cassini, wu2024falcon}.

\begin{figure}[tb]
    \centering
    \includegraphics[width=\linewidth]{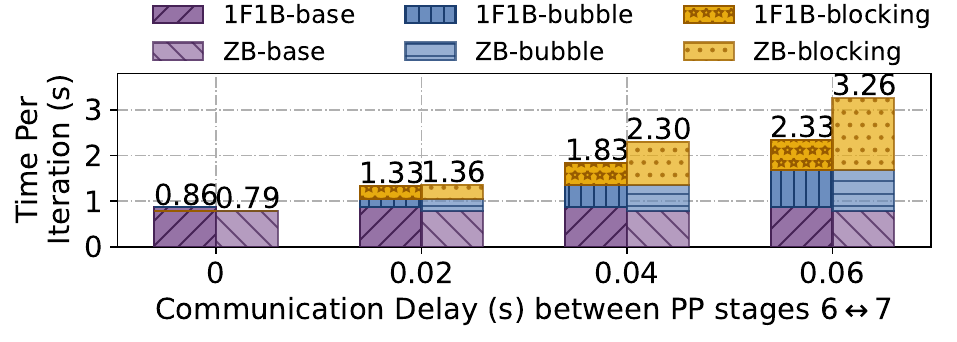}
    \vspace{-0.5cm}
    \caption{GPT-2 14B training performance on 8 nodes (one H800 GPU per node) with 8-stage PP, where minor communication delays between PP stages trigger dependency bubbles and blocking stalls, causing significant slowdowns in 1F1B and ZeroBubble (ZB) pipelines.}
    \label{fig:intro_problems}
\end{figure}

Production systems mainly focus on detecting communication stragglers in large-scale
training~\cite{cao2024crux, dubey2024llama, wu2024falcon, cui2025xputimer}
and rely on traffic load balancing at flow or packet level to alleviate network
congestion~\cite{le2024strack, gangidi2024rdma, dixit2013impact,
alizadeh2014conga}. However, these approaches are \emph{agnostic} to the parallelism
strategies of the training job and cannot effectively mitigate the
straggler impacts on job performance. As illustrated in \autoref
{fig:intro_problems}, with pipeline parallelism (PP), even minor
communication delays between two PP stages can result in significant training
slowdowns, which grow rapidly as the delay increases.
Our study identifies two issues that contribute to this large slowdown, with
their impacts shown in \autoref{fig:intro_problems}.

First, given sophisticated data dependencies in
a PP schedule (e.g., Gpipe~\cite{huang2019gpipe}, 1F1B~\cite
{narayanan2019pipedream} and ZeroBubble~\cite{qi2023zero}), a single
communication straggler, when exceeding a certain threshold, can set off a
domino effect, triggering \emph{cascading bubbles} (i.e., GPU idle periods) that
propagate across subsequent stages (\autoref{fig:delay_zb}-bottom). These
bubbles, which we call \emph{dependency bubbles}, cause severe misalignment
within the pipeline, disrupting its entire schedule.

\begin{figure}
    \centering
    \includegraphics[width=\linewidth]{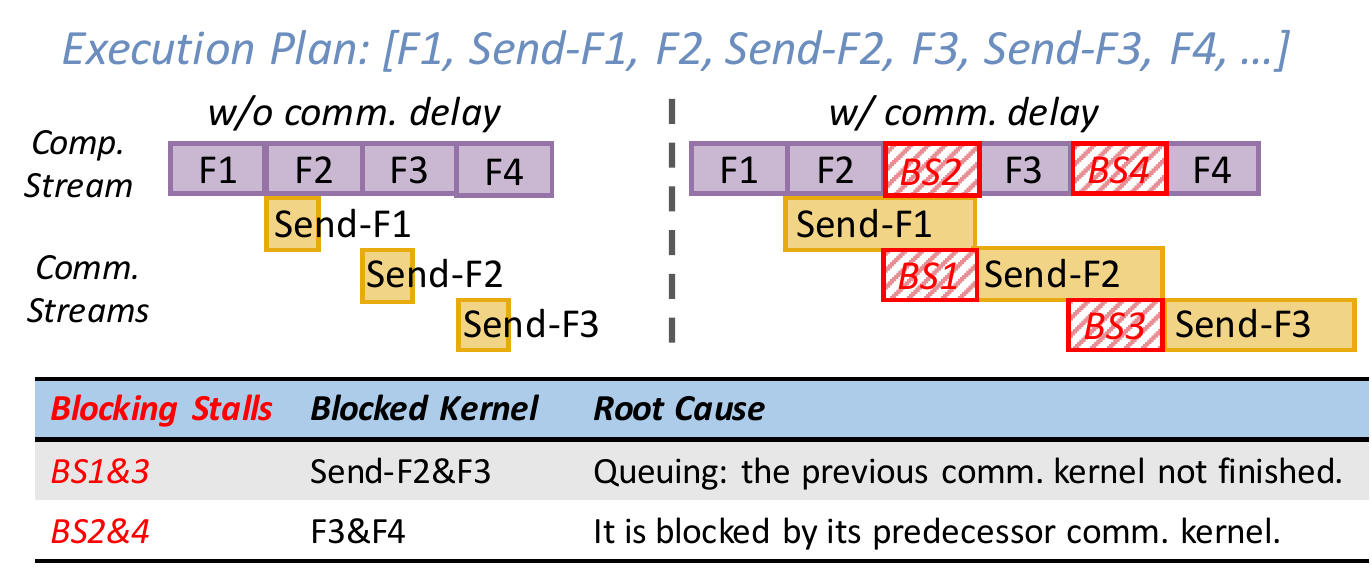}
    \vspace{-0.5cm}
    \caption{Head-of-line blocking due to sequential kernel scheduling: slow comm.~blocks subsequent comp.}
    \label{fig:intro_kernel}
\end{figure}

Second, existing GPU kernel scheduling is susceptible to \emph
{head-of-line blocking}, where slow communication can block subsequent
computation. To illustrate this, we refer to \autoref{fig:intro_kernel}. 
Once a PP schedule is determined, a low-level kernel execution plan is
generated accordingly, in which communication and computation operations are
interleaved to overlap communication latency (e.g., \texttt{F1}, \texttt
{Send-F1}, \texttt{F2}, \texttt{Send-F2}, $\dots$). The GPU scheduler \emph
{sequentially schedules} operations following this plan.
However, in the presence of a slow link, a communication operation cannot be
scheduled immediately as previous communication operations are
still queued up, pending for transmission. This blocks the
scheduling of subsequent computation operations, introducing additional \emph
{blocking stalls} to the pipeline (\autoref{fig:intro_kernel}-right),
which in turn triggers more bubbles, further
aggravating training slowdowns.


Eliminating dependency bubbles and blocking stalls requires dynamically
adapting the pipeline schedule with framework support, which remains lacking
in existing systems. For instance, \GreyHound~\cite{wu2024falcon} simply
reassigns slow links from communication-heavy DP groups to communication-light PP groups,
without addressing subsequent pipeline bubbles. Recycle~\cite
{gandhi2024recycle} and Oobleck~\cite{oobleck} pre-compute a pipeline
reconfiguration plan for handling GPU failures. However, communication
stragglers cannot be addressed using a pre-computed plan as pipeline
adaptation must be made dynamically based on the changing straggler magnitude.


In this paper, we present \SystemName{}, a straggler-resilient system for
efficient hybrid-parallel training. It addresses dependency bubbles and
blocking stalls caused by slow communications with two key designs.


\PHM{Straggler-resilient pipeline adaptation.} 
We show analytically that a pipeline schedule (e.g., 1F1B~\cite
{narayanan2019pipedream} or ZeroBubble~\cite{qi2023zero}) can tolerate slow
communication up to a certain threshold without triggering cascading bubbles
in a domino effect. This threshold is in proportion to the \emph
{slackness} between adjacent PP stages, defined as the \emph{difference of
the number of forward operations} scheduled in the two stages during the warm-up
phase. Larger slackness enhances the pipeline's resilience to a longer
communication delay. Based on this, \SystemName{} initially generates a
ZeroBubble schedule~\cite{qi2023zero} that maximizes the minimum inter-stage slackness under
memory and configuration constraints. It then monitors communication delays
between PP stages. When the delay exceeds the tolerance threshold (given by
our analytical model), it quickly reacts by adapting the pipeline schedule to
increase the inter-stage slackness of the slow link, eliminating all or most
straggler-induced dependency bubbles.


\PHM{Decoupled data plane.} 
\SystemName further employs a \emph{fully-decoupled data plane} to address
head-of-line blocking of GPU kernel scheduling and the resulting blocking
stalls. Upon detecting communication stragglers, the system transparently
switches to a \emph{delegation mode}, in which it offloads PP communications
from GPU to host memory and uses dedicated delegate processes to perform data
transfer via CPU-side RDMA. \SystemName{} chooses to bypass the more
efficient GPU-direct RDMA due to three design imperatives. First, it
completely decouples PP communications from GPU execution, preventing slow
communication from blocking subsequent GPU computations. Second, optimally
adapting pipeline schedule in the presence of stragglers may require storing
more activations than GPU memory can hold, where offloading to host memory
becomes necessary. Third, given that PP schedule only requires light to
moderate communications (compared to DP and TP), the performance overhead
introduced by offloading can be minimized with system-level optimizations. This design additionally enables RNIC
fault tolerance: upon detecting a RNIC failure, the system reroutes traffics
through remaining healthy RNICs via the delegation path, obviating the need for
checkpoint-and-restart failovers.

We implemented \SystemName on top of the Megatron-LM~\cite
{shoeybi2019megatron, narayanan2021megatron} training framework, using
ZeroBubble~\cite{qi2023zero} as the base pipeline schedule to achieve the
best performance. We evaluated \SystemName using GPT-2 models of varying sizes,
from 7B to 140B parameters, on H800 clusters. Compared to state-of-the-art
straggler mitigation solutions, \SystemName reduces the average training
iteration time by 1.2-3.5$\times$ under various network latency conditions.
In a large-scale deployment involving 128 H800 GPUs, \SystemName consistently
delivers high training throughput in the presence of frequent communication
stragglers, outperforming the baselines by 1.41$\times$ while resilient to
RNIC failures.


\section{Background and Motivation}
\label{sec:background}

\subsection{Hybrid-Parallel DNN Training}
The increasing scale of Deep Neural Networks (DNNs), driven by empirical
scaling laws~\cite{kaplan2020scalinglaws}, has led to state-of-the-art
models with hundreds of billions of parameters~\cite
{dubey2024llama,zhang2022opt,achiam2023gpt,bloom176b, team2023gemini,
liu2024deepseek, guo2025deepseek}. Training such large models requires
high-performance computing (HPC) clusters comprising tens of thousands of
GPUs~\cite{jiang2024megascale,qian2024alibaba,narayanan2021megatron}, leveraging
advanced parallelization strategies in three primary forms.

\PHM{Data parallelism (DP)} distributes identical model replicas across GPU
 groups, with each replica processing a subset of the input data (mini-batches)
 concurrently~\cite
 {rajbhandari2020zero,shoeybi2019megatron,narayanan2021megatron}.
 Synchronization is performed at the end of each iteration via all-reduce operations,
 often spanning multiple GPU nodes interconnected through high-speed networks such as RDMA over Infiniband (IB) ~\cite{gangidi2024rdma,qian2024alibaba,jiang2024megascale}.

\PHM{Tensor parallelism (TP)} partitions individual tensors (e.g., weight
 matrices) within model layers across multiple GPUs~\cite
 {zheng2022alpa,shoeybi2019megatron}. While this technique parallelizes linear
 algebra operations within layers, it incurs significant communication overhead
 due to frequent reduce-scatter and all-reduce operations during forward and
 backward passes. Consequently, TP is typically restricted to 
 single-node deployments with high-bandwidth GPU interconnects (e.g., NVLink).




\PHM{Pipeline parallelism (PP)} divides the model into sequential layer
 groups (\emph{stages}) assigned to different
 GPUs~\cite{huang2019gpipe,zheng2022alpa,narayanan2021megatron}. 
 Mini-batches are further split into micro-batches that flow through these stages in a
 pipelined manner, enabling parallel processing across stages.
 PP requires lower network bandwidth than DP/TP
 by communicating only activations and gradients at layer boundaries.
 However, it can suffer from reduced training throughput due to pipeline
 dependencies and bubbles (idle slots) as the number of
 stages increases. Modern PP implementations like 1F1B~\cite
 {narayanan2019pipedream} and ZeroBubble (ZB)~\cite{qi2023zero} address these
 issues, with ZB achieving bubble-free execution at the expense of increased memory footprint.

\subsection{Reliability Issues}

Training large DNN models over extended periods face
reliability challenges, primarily manifested through \emph{crash failures} and
still-functioning but slow \emph{stragglers}~\cite
{dubey2024llama,zhang2022opt,wu2024falcon,jiang2024megascale,oobleck,wang2023gemini,cui2025xputimer}.
These issues stem from hardware failures, software errors, or resource
contention, with even a single affected component disrupting the entire
distributed training.

\PHM{Crash failures.} 
Crash failures have been extensively analyzed in recent studies~\cite
{jeon2019analysis,hu2024characterization,xiong2024superbench,dubey2024llama,mohan2021checkfreq,wang2023gemini,thorpe2023bamboo,oobleck,gandhi2024recycle}.
Meta and ByteDance report that faulty GPUs and RNICs are the leading causes,
accounting for 40\% and 10\% of failures, respectively~\cite
{dubey2024llama,hu2024characterization}. To address this, researchers have
developed \emph{fault-resilient solutions} for hybrid-parallel training,
including (1) optimized checkpoint-and-restart mechanisms to minimize recovery overhead~\cite
{mohan2021checkfreq,wang2023gemini} and (2) exploiting PP's 
functional redundancy to reduce checkpointing~\cite
{oobleck,thorpe2023bamboo,gandhi2024recycle}.

\PHM{Stragglers.} 
Stragglers—manifested as slow computation or communication—represent another major
reliability concern in large-scale DNN training~\cite{wu2024falcon,cui2025xputimer,dubey2024llama}. Prior
studies~\cite{wu2024falcon,jiang2024megascale} indicate that \emph
{computation stragglers}, typically caused by GPU thermal throttling or
hardware defects, are relatively rare ($\sim$0.5\% occurrence) and short-living
(usually recovering in 10 minutes). These incidents can be efficiently addressed by
adjusting the parallelization of GPU devices~\cite{li2024malleus,wu2024falcon}. 
In contrast, \emph{communication stragglers}, predominantly due to network
congestion, are more frequent and persistent, often lasting for hours~\cite
{cao2024crux,wu2024falcon}. In Alibaba's production multi-tenant clusters, over 36\%
of jobs using >50\% GPUs experience slowdowns from communication
issues~\cite{cao2024crux}. In extreme cases, such stragglers can reduce 
training throughput by up to 90\%~\cite{wu2024falcon}.

\begin{figure}[tb]
    \centering
    \includegraphics[width=\linewidth]{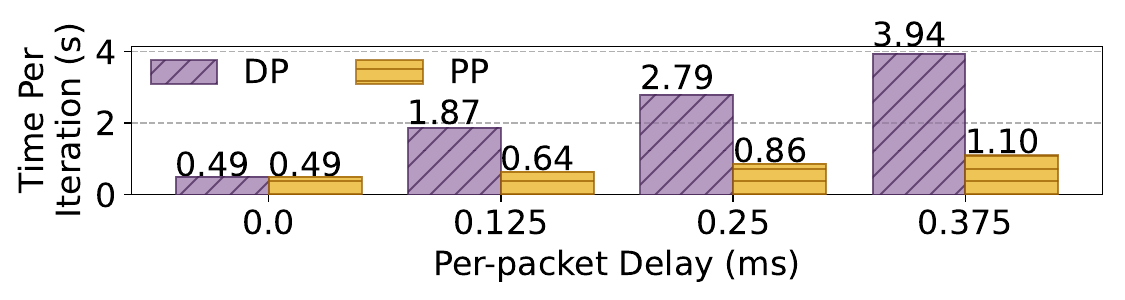}
    \caption{Iteration time growth under different per packet delays, where DP is more communication-sensitive to PP.}
    \label{fig:motivation_dp}
\end{figure}

Communication stragglers may occur on cross-node links between DP groups
(DP straggler) or between PP stages (PP straggler), but not in TP groups, as TP
communications are confined to intra-node, high-bandwidth, and stable 
NVLinks~\cite
{wu2024falcon,cao2024crux}. Compared to PP stragglers, DP stragglers can have
a more pronounced impact on performance, as DP's all-reduce operations incur
substantial communication costs and are bottlenecked by the slowest link in
the all-reduce ring~\cite{chen2016revisiting, harlap2016addressing,
zhou2020falcon, wu2024falcon, li2024malleus}. 

To illustrate this effect, we train a GPT2-7B model on 8 nodes (each with one
H800 GPU) by configuring a (2 DP, 4 PP) ZeroBubble pipeline using
Megatron-LM~\cite{narayanan2021megatron}. We manually inject a per-packet
delay of 0.125/0.25/0.375~ms (corresponding to 10/20/30 ms inter-PP stage
latency) into a designated cross-node link (400 Gbps IB). As shown in \autoref
{fig:motivation_dp}, when this link is part of the DP communication group,
iteration time increases by up to $8.04\times$. In contrast, when the
same link is assigned for PP communication, the
slowdown is limited to $\le 2.24\times$. Motivated by these findings, recent
work~\cite{wu2024falcon} proposes mitigating DP stragglers by reconfiguring
the parallelization scheme to assign slow links to PP stages instead of DP
groups. While this approach effectively converts a DP straggler to a PP
straggler, it still results in significant slowdowns, which remain open to
address.

\section{Impact Analysis and Challenges}
\label{sec:challenges}



\begin{figure}[tb]
    \centering
    \includegraphics[width=0.95\linewidth]{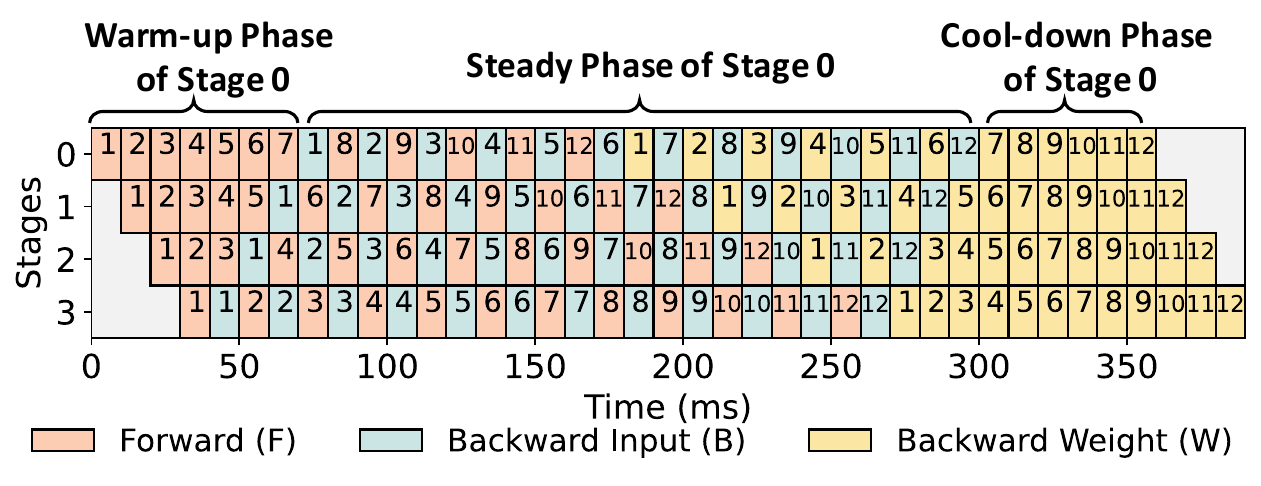}
    \caption{An ideal straggler-free ZeroBubble~\cite{qi2023zero} pipeline with $4$ stages and $12$ microbatches, completing in 390 ms.}
    \label{fig:ideal_zb}
\end{figure}

In this section, we systematically investigate how inter-stage communication
stragglers lead to substantial pipeline stalls. Our investigation focuses
on ZeroBubble (ZB) pipeline scheduling~\cite{qi2023zero}---a generalized,
fine-grained PP scheduling scheme that subsumes common approaches like
1F1B~\cite{narayanan2019pipedream} as special cases. As illustrated
in \autoref{fig:ideal_zb}, ZB eliminates pipeline bubbles by decomposing
backward pass into two independent operators, backward input ($B$) and backward
weight ($W$), then precisely orchestrating their execution. This generalized
design encapsulates diverse pipeline behaviors, ensuring the broad
applicability of our findings. In the following, we identify two critical
delay propagation mechanisms: (1) domino effect of cascading bubbles due to
PP's vulnerable dependency chains (\S\ref{sec:bubbles}) and (2) head-of-line
blocking stalls due to sequential GPU kernel scheduling (\S\ref
{sec:interference}).


\subsection{Domino Effect of Cascading Bubbles}
\label{sec:bubbles}

Pipeline parallelism orchestrates stage execution with strict data
dependencies. In ZB scheduling, these dependencies manifest through two key
constraints: (1) a forward operator ($F_j$) in stage $S_i$ requires
completion of $F_j$ in preceding stage $S_{i-1}$; (2) backward operators
$B_j$ and $W_j$ in $S_i$ must be scheduled after $B_j$ completion in
subsequent stage $S_{i+1}$. The presence of communication stragglers between
stages $S_i$ and $S_{i+1}$ introduces additional latency $c_i$, with two
immediate impacts: (1) forward operator $F_j$ in stage $S_{i+1}$ can only
commence after the completion of $F_j$ in $S_i$ plus the communication delay
$c_i$; (2) backward operators $B_j$ and $W_j$ in $S_i$ must follow the
completion of $B_j$ in $S_{i+1}$ plus the delay $c_i$.

\begin{figure}[tb]
    \centering
    \includegraphics[width=0.95\linewidth]{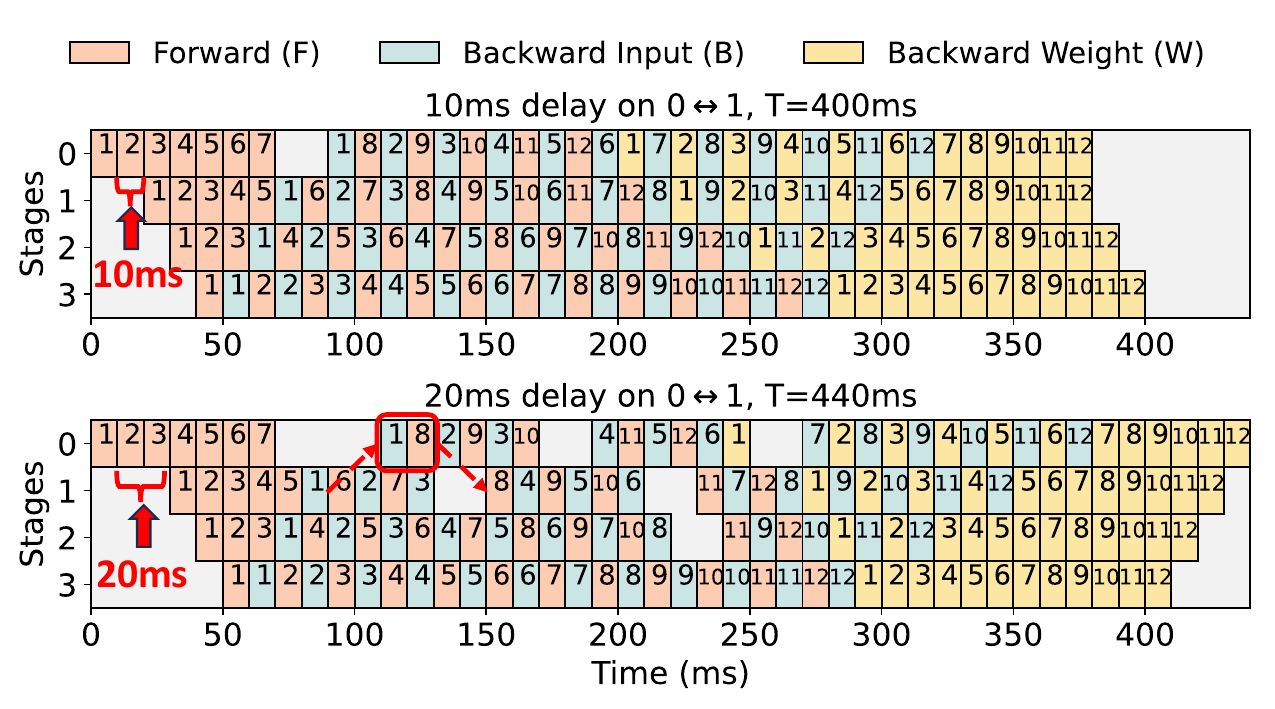}
    \caption{ZeroBubble schedule under $c_0=10/20$ ms delay between stage 0 and 1. Increasing $c_0$ from 0 to 10 ms only prolongs iteration time $T$ by 10 ms, while an additional 10 ms delay introduces a 40 ms growth in $T$.}
    \label{fig:delay_zb}
\end{figure}

We quantify the straggler impacts on pipeline schedules through simulations of
a 4-stage pipeline processing 12 microbatches, with uniform operator
execution time ($F=B=W=10$ ms). \autoref{fig:ideal_zb} shows the
ideal straggler-free ZB schedule completing in 390 ms with zero bubbles. 
\autoref{fig:delay_zb} depicts the resulting schedules in the presence of
 communication delays of 10 and 20~ms between stages 0 and 1. A 10~ms delay
 introduces a linear slowdown, extending the execution time from 390 to 400
 ms. However, further increasing the delay to 20~ms results in a \emph
 {non-linear growth} of pipeline stall to 440 ms. This occurs because the
 20~ms delay pushes the first backward $B_1$ and subsequent $F_8$ in $S_0$
 back to $t=110$~ms, creating a bubble in $S_1$. This bubble propagates
 downstream, triggering cascading pipeline stalls in subsequent stages.

Extending our analysis, \autoref{fig:simdelay} examines various ZB pipelines
with different slackness parameters $\Delta_i$ (defined in \S\ref
{sec:analysis}). For each pipeline schedule, we gradually increase the
communication delays between adjacent stages and depict in \autoref{fig:simdelay} the
resulting pipeline delays (left) and bubble rates (right). These results empirically
demonstrate that localized communication delays exceeding a certain
threshold create cascading dependency bubbles in a domino effect, 
leading to significant global pipeline stalls.




\subsection{Head-of-Line Blocking Stalls}
\label{sec:interference}

\begin{figure}[tb]
    \centering
    \includegraphics[width=0.95\linewidth]{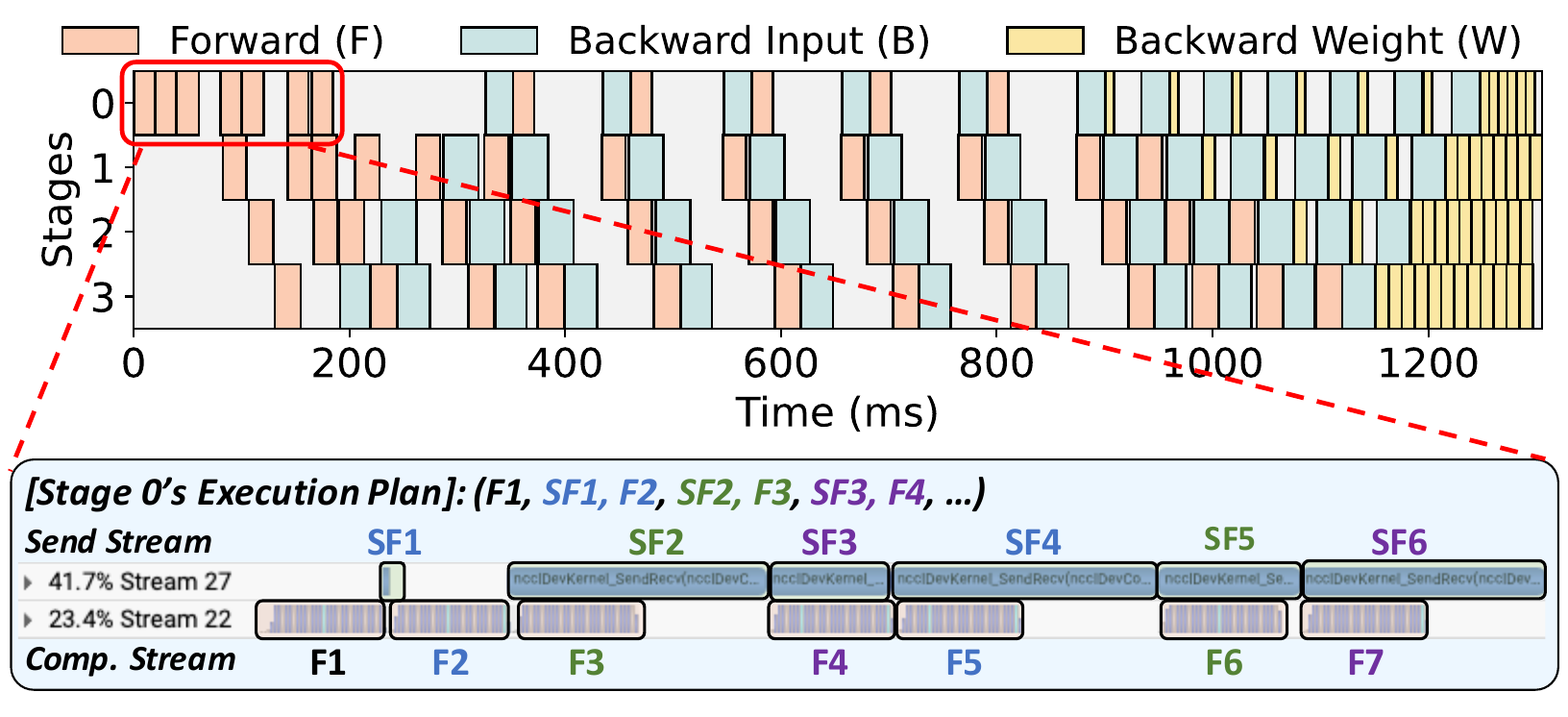}
    \caption{Slow communication ($SF_2$) induces HOL blocking stalls ($F_4$) due to sequential GPU kernel scheduling.}
    \label{fig:profile_zb}
\end{figure}

Communication stragglers further degrade pipeline performance through
low-level GPU kernel scheduling anomalies, manifested as \emph{head-of-line
blocking stalls}. To demonstrate this problem, we conduct a GPT2-7B
training experiment across four nodes (one H800 GPU per node), under a ZB
schedule of(1 TP, 1 DP, 4 PP). We inject a 30~ms delay between PP stages 0
and 1 using NCCL network plugin. \autoref{fig:profile_zb} illustrates the
profiled kernel execution result using NVIDIA Nsight Systems~\cite{nsys}.


\PHM{Blocking stalls.} 
Given a pipeline schedule, a training framework (e.g., Megatron-LM~\cite{narayanan2021megatron},
TorchTitan~\cite{liang2024torchtitan}) generates a \emph{fixed execution plan} that
interleaves computation ($F$, $B$, $W$) and communication operations, including
send/recv-forward ($SF$/$RF$) and send/recv-backward ($SB$/$RB$). 
This execution plan maximizes computation-communication overlap 
through a carefully ordered operation scheduling sequence (e.g., [$F_1, SF_1,
F_2, SF_2$, $\dots$] in \autoref{fig:profile_zb}). The kernel scheduler
sequentially launches these operations following this predetermined order.
However, delayed communication operations stall subsequent computation
operations, creating unexpected bubbles. In
\autoref{fig:profile_zb}, the delayed $SF_2$ blocks the launching of $F_4$
in stage~0, despite it being ready to execute after $F_3$.

\PHM{Root cause analysis.} 
Blocking stalls arise when NCCL's transmission queue fills, which may
disrupt CUDA's asynchronous execution. This is because each NCCL \texttt
{send}/\texttt{recv} launches a GPU kernel to enqueue data transfers, which
are handled asynchronously by a dedicated backend thread. Under slow links,
the queue builds up as pending transfers outpace actual data transmission.
The launching of subsequent communication kernels (e.g., $SF_3$) hence
blocks—they do not return control to the kernel scheduler until the queue
space becomes available. This in turn prevents the CUDA scheduler from
launching following computation kernels (e.g., $F_4$ is blocked until
$SF_3$ returns control, which itself waits on $SF_2$), thus creating
head-of-line (HOL) blocking stalls in computation. Notably, CUDA
multi-streaming fails to mitigate this issue because all streams within a GPU
context share the same NCCL communicator and transmission queue.

\subsection{Which Layer to Optimize?}
\label{sec:which_layer_to_optimize}

Our analysis reveals that communication stragglers degrade pipeline
performance through two mechanisms: dependency bubbles and HOL blocking
stalls (see \autoref{fig:intro_problems} for quantitative contributions).
Addressing these issues requires careful considerations of optimization
layers within the system stack.

\textbf{Network-level optimizations}, such as
ECMP~\cite{gangidi2024rdma, thaler2000multipath, alizadeh2014conga} or packet
spraying~\cite{le2024strack,dixit2013impact}, balance traffic at the flow or
packet level. While effective for general congestion reduction, these
approaches are \emph{agnostic} to training semantics (e.g., parallelism
strategies and communication patterns) and cannot prioritize critical
communications over less sensitive transfers, nor can they address HOL blocking
stalls. Also, even state-of-the-art load balancing techniques
cannot eliminate network congestion, especially in multi-tenant
clusters~\cite{cao2024crux, addanki2024challenging}. 


Effective straggler mitigation requires \textbf
{framework-level optimizations}, leveraging training semantics such
as pipeline schedule, operator dependencies, and communication patterns. This
semantic insight enables targeted mitigation strategies, such as pipeline
adaptation and blocking-free kernel scheduling---none of these can be
implemented in the network layer. 
However, existing reliability enhancement mechanisms for training frameworks
are ineffective in addressing communication stragglers under pipeline parallelism.
For instance, Malleus~\cite{li2024malleus} exclusively targets computation stragglers
without considering slow communications. XPUTimer~\cite{cui2025xputimer} 
provides production-grade straggler detection yet lacks integrated mitigation
mechanisms. \GreyHound~\cite{wu2024falcon} reassigns slow links from DP groups
to PP stages, but fails to resolve resulting PP stragglers. Recycle~\cite
{gandhi2024recycle} and Oobleck~\cite{oobleck} rely on static pipeline 
reconfiguration plans to handle GPU failures, lacking adaptability to dynamic 
network conditions.

\section{Straggler-Resilient Pipeline Adaptation}
\label{sec:algorithm_design}

\SystemName{} is a system that effectively mitigates communication stragglers
 for hybrid-parallel training with two key designs: (1) a straggler-resilient
 pipeline adaptation algorithm that dynamically adapts the pipeline schedule
 to minimize dependency bubbles (\S\ref{sec:bubbles}), and (2) a
 fully-decoupled data plane eliminating HOL blocking stalls (\S\ref{sec:interference}).

In this section, we describe the first design component, the pipeline
adaptation algorithm, where \emph{we assume no HOL blocking stalls}---which is
guaranteed by our second design in \S\ref{sec:system_design}. We first
analytically quantify the accumulated pipeline delays caused by a slow link
between PP stages (\S\ref{sec:analysis}). Driven by this analytical result,
we design the pipeline adaptation algorithm, including warm-up scheduling
(\S\ref{sec:initpipe}) and full pipeline scheduling (\S\ref
{sec:pipeadapt}). Key mathematical notations are summarized in
\autoref{tab:notations} to guide subsequent analysis.

\subsection{Quantitative Delay Analysis}
\label{sec:analysis}

\begin{table}[tb]
    \centering
    \footnotesize
    \begin{tabular}{c|c}
        \hline
        \textbf{Notation} & \textbf{Explanation} \\
        \hline
        $S_i$ & The $i$\textsuperscript{th} pipeline stage.\\
        $S$ & Total number of pipeline stages. \\
        $N$ & Total number of microbatches.\\
        $t$ & Execution time of a single operation $F/B/W$.\\
        $c_i$ & Communication latency between $S_i$ and $S_{i+1}$.\\
        $T$ & The overall pipeline execution time.\\
        $x_i$ & Number of warm-up forwards in stage $S_i$.\\
        $\Delta_i$ & Slackness between stages $S_i$ and $S_{i+1}$.\\
        $\delta$ & Simulator's time step size.\\
        \hline
    \end{tabular}
    \caption{Notations in the quantitative analysis and algorithms.}
    \label{tab:notations}
\end{table}

\PHB{Key insight.}
In \S\ref{sec:bubbles}, we empirically demonstrate that communication delays
exceeding a certain threshold induce disproportionately significant pipeline
stalls through cascading bubbles. We further develop an analytical model to
quantify this effect. Our analysis identifies the \emph{slackness} of a pipeline
as the key structural resilience to communication delays, informally defined
as \emph{the difference of the warm-up forward counts in two adjacent stages}.
Intuitively, the more warm-up forward operators the pipeline schedules in stage $S_i$
than in $S_{i+1}$, the larger slackness it provides between
the two stages, which can be utilized to ``absorb'' more dependency bubbles
caused by inter-stage communication delays.

\PHM{Analysis.}
To prove this result, we base our analysis on ZeroBubble (ZB) pipeline
scheduling, as it is a more generalized design encapsulating common
approaches like 1F1B as special cases without backward weight ($W$) costs.
Our analytical findings are hence broadly applicable to 1F1B and other
pipeline scheduling approaches. 

In a ZB schedule, each pipeline stage operates through three phases (\autoref
{fig:ideal_zb}): (1) the \textbf{warm-up phase} containing a configurable
number of forward-only ($F$) operations, (2) the \textbf{steady phase}
containing a mixture of forward ($F$), backward input ($B$), and backward
weight ($W$) operations, and finally (3) the \textbf{cool-down phase}
containing the remaining backward weight ($W$) computations. For ease of
presentation, we assume all three operations $F$, $B$ and $W$ have a uniform execution time 
($t^F = t^B = t^W = t$). Nonetheless, our analysis extends to a more general
heterogeneous setting. Let $x_i$ be the number of forward operations scheduled
in stage $S_i$ during the warm-up phase, aka \emph{warm-up forward count}.
The following lemma shows that $x_i$ is monotonically decreasing, with the proof
given in the Appendix.






\begin{lemma}[Monotonic warm-up]
\label{lem:non_increasing}
For any pipeline schedule, the warm-up forward count is non-increasing over stages, 
i.e., $x_i \geq x_{i+1}$ for all $i = 0, 1, \dots, S-1$.
\end{lemma}

We formally define the slackness between stages $S_i$ and $S_{i+1}$ as
$\Delta_i = x_i - x_{i+1}$, which is guaranteed non-negative
by Lemma~\ref{lem:non_increasing}. The following theorem identifies
the slackness as the key structural resilience to communication delays.


\begin{theorem}[Delay resilience] 
\label{thm:resilience}
  Let $c_i$ be the communication delay between stages $S_i$ and $S_{i+1}$.
  The accumulated pipeline delay caused by $c_i$ is $\Theta(c_i)$ if $c_i \le (\Delta_i-1)t$ 
  but amplifies to $\Theta\left(\frac{N c_i}{\Delta_i + 1}\right)$ if $c_i > (\Delta_i-1)t$,
  where $t$ is the operation execution time (i.e., $t^F = t^B = t^W = t$).
\end{theorem}

\begin{figure}[tb]
    \centering
    \includegraphics[width=0.95\linewidth]{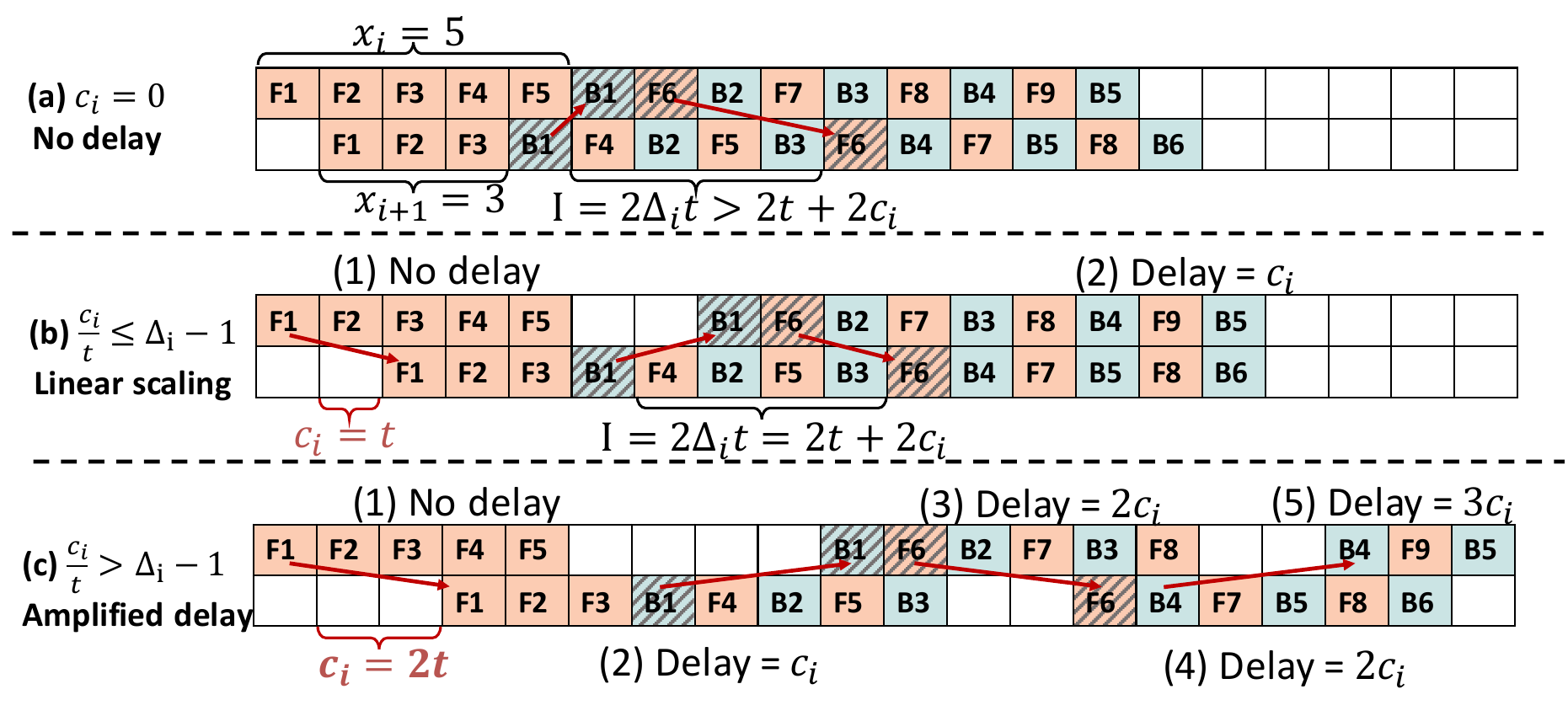}
    \caption{Analysis of accumulated pipeline delay with $\Delta_i=2$.}
    \label{fig:delay_analysis}
\end{figure}

\textbf{Proof:} By the Lemma, $\Delta_i$ is non-negative, so we can prove this theorem by considering the following two cases.

\underline{\textit{Case 1: $c_i \le (\Delta_i-1)t$.}} For any two \textit{adjacent} $B, F$ operations $B_{i,a}, F_{i,b}$ in $S_i$ (e.g., $B_1, F_6$ in the figure), we can find its corresponding operations $B_{i+1,a}, F_{i+1, b}$ in $S_{i+1}$, and define the \emph{feasible interval} $I_i$ as the interval between the end of $B_{i+1,a}$ and the start of $F_{i+1,b}$. During the steady phase, this interval is inherently $2\Delta_i t$ in an ideal no-delay scenario (\autoref{fig:delay_analysis} (a)). To absorb the communication delay, the total execution time of the following operations must not exceed $I_i$: (1) sending $B_{i,a}$ back to $S_i$, (2) calculation of $B_{i,a}$ and $F_{i,b}$ in $S_i$, and (3) sending $F_{i,b}$ to $S_{i+1}$ (costs $2c_i+2t$ in total).

Therefore, $c_i \le (\Delta_i - 1)t$ ensures that the $2\Delta_i t$ interval is larger than the $2c_i+2t$ cost. Thus, the delay $c_i$ is fully absorbed without propagating bubbles to subsequent operations (\autoref{fig:delay_analysis} (b)). The accumulated delay is therefore bounded by $\Theta(c_i)$, as no cascading stalls occur.

\underline{\textit{Case 2:} $c_i > (\Delta_i-1)t$}. For this case, the pipeline incurs cascading bubbles as illustrated in \autoref{fig:delay_analysis} (c). This is due to each feasible interval $I$ should be expanded from $2\Delta_i$ to $2c_i+2t$ to fit the communication and computation operations. This expansion postpones a group of $2\Delta_i+2$ operations by $2c_i$, and the delay will accumulate to the subsequent group of operations. Therefore, for a pipeline with $N$ operations, the overall delay will be $
\Theta\left(\frac{N c_i}{\Delta_i + 1}\right)$, as each group of $\Delta_i + 1$ operations contributes $c_i$ to the total. \unskip\nobreak\hfill $\square$

Theorem~\ref{thm:resilience} essentially states that a pipeline with slackness
$\Delta_i$ can tolerate a communication delay up to $(\Delta_i-1)t$ without
triggering cascading bubbles. This result can be extended to a more general
setting where the execution times of $F$, $B$ and $W$ are non-uniform. 
In this case, communication delay $c_i$ will not introduce cascading
bubbles \emph{if and only if} 
\begin{equation}
\label{eq:absorb}
    \begin{aligned}
    t^F_i + t^B_i + 2c_i \le \Delta_i (t^F_{i+1}+ t^B_{i+1}),
    \end{aligned}
\end{equation}
where $t^F_{i}$ and $t^B_{i}$ denote the execution time of forward $F$ and
backward input $B$ operations in stage $i$, respectively. When there are
multiple slow links, the accumulated delay is simply the summation of all
stragglers' individual contributions.

\begin{figure}
    \centering
    \includegraphics[width=0.95\linewidth]{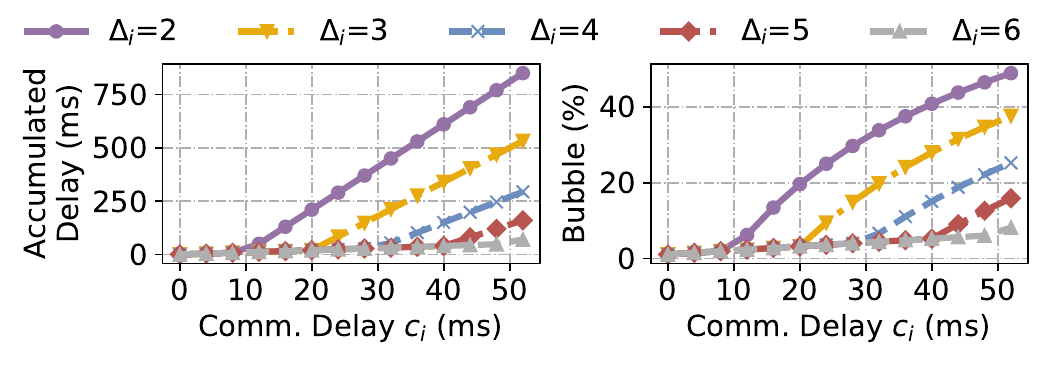}
    \caption{Simulated delay and bubble rate using different $\Delta_i$ for a pipeline with $N=30$ microbatches and $t=10$ ms.}
    \label{fig:simdelay}
\end{figure}

\autoref{fig:simdelay} empirically verifies our analytical findings in
 simulations: as the communication delay $c_i$ grows beyond the threshold $
 (\Delta_i - 1)t$, the accumulated pipeline delay sharply increases, 
 aligning with that predicted by Theorem~\ref{thm:resilience}.

\subsection{Orchestrating Warm-up Forwards}
\label{sec:initpipe}

Our previous analysis indicates that enhancing the pipeline's resilience to
communication delays requires configuring a larger slackness between two
stages. However, doing this comes at a cost of increased memory footprint, as
more forward activations are maintained on device. We design pipeline
scheduling algorithms that optimally orchestrate warm-up forwards in each
stage (i.e., $x_i$), maximizing the straggler resilience under the memory
constraint. Our algorithms include two strategies: (1) \emph
{initial planning} for pipeline initialization and (2) \emph
{dynamic adaptation} for reconfiguring the pipeline in response to straggler
presence. We next explain how the two strategies orchestrate warm-up forwards
in each stage, followed by constructing the full-pipeline schedule in \S\ref
{sec:pipeadapt}.



\begin{algorithm}[tb]
\small
\caption{Initial Planning}
\label{alg:warmup_fwds}
\begin{algorithmic}[1]
\Require Number of pipeline stages $S$, available GPU memory capacity $M$, per-activation memory $M^F$.
\Ensure Initial warm-up forward counts $\{x_0,\ldots, x_{S-1}\}$.
\Function{GetInitWarmupFwds}{$S, M, M^F$}
    \State $x_{\max} = \lfloor \frac{M}{M^F}\rfloor$ \Comment{\textit{\textcolor{blue}{Calculate max \#activations on GPU.}}} \label{line:max_fwd}
    \State $x_0 \gets x_{\max}$ \Comment{\textit{\textcolor{blue}{First stage holds $x_{\max}$ forwards.}}} \label{line:x_0}
    \State $\Delta_{\text{avg}} \gets \lfloor \frac{x_{\max} - 1}{S - 1}\rfloor$ \label{line:avg_gap}
    \State $r \gets (x_{\max} - 1) \bmod (S - 1)$
    \For{$i \gets 1$ \textbf{to} $S-1$}
        \State $\triangleright$ \textit{\textcolor{blue}{Calculate $\Delta_{i-1}$ and $x_i$ recursively.}}
        \State $\Delta_{i-1} \gets \Delta_{\text{avg}} + 1$ \textbf{ if } $i \le r$ \textbf{ else } $\Delta_{\text{avg}}$
        \State $x_i \gets x_{i-1} - \Delta_{i-1}$ \label{line:x_i}
    \EndFor
    \State \Return $\{x_0,\ldots, x_{S-1}\}$
\EndFunction
\end{algorithmic}
\end{algorithm}

\PHM{Initial planning.} 
During pipeline initialization, the system assumes no knowledge of stragglers.
As they can occur on any link between two stages, the best strategy is to \emph
{maximize the minimum inter-stage slackness} within the pipeline, i.e., $\max \min_i \Delta_i$. This is equivalent to configuring a pipeline
that \emph{uniformly maximizes each} $\Delta_i$ under GPU memory
constraints. \autoref{alg:warmup_fwds} shows how this can be achieved. It
first computes the maximum number of forward activations that a GPU can
maintain in memory, all of which are assigned to stage 0 
(lines~\ref{line:max_fwd} and \ref{line:x_0}). With $x_0$ determined, it
then computes the warm-up forward counts in subsequent stages to ensure that
$x_i$'s are monotonically decreasing (Lemma~\ref{lem:non_increasing}) with 
as balanced slackness $\Delta_i$ as possible (lines~\ref{line:avg_gap} to \ref{line:x_i}). 
The generated pipeline provides the maximum uniform delay resilience.


\begin{algorithm}[tb]
\small
\caption{Dynamic Adaptation}
\label{alg:adapt_fwds}
\begin{algorithmic}[1]
\Require Number of pipeline stages $S$, per-stage forward/backward times $\{t^{F}_i\}, \{t^B_i\}$, inter-stage communication delays $\{c_i\}$.
\Ensure Adapted warm-up forward counts $\{x_0,\ldots, x_{S-1}\}$.
\Function{GetAdaptedWarmupFwds}{$S, \{t^{F}_i\}, \{t^B_i\}, \{c_i\}$}
    \State $\triangleright$ \textit{\textcolor{blue}{The last stage holds only one warm-up forward.}}
    \State $x_{S-1} \gets 1$ \label{line:last_stage}
    \State $\triangleright$ \textit{\textcolor{blue}{Calculate $\Delta_i$ and $x_i$ backward-recursively.}}
    \For{$i \gets S-2$ \textbf{to} $0$}
        \State $\triangleright$ \textit{\textcolor{blue}{Ensure bubble absorption by \autoref{eq:absorb}.}}
        \State $\Delta_i \gets \min\left(N-2S, \max\left( \left\lceil\frac{t^F_i+t^B_i+2c_i}{t^F_{i+1}+t^B_{i+1}}\right\rceil, 2\right)\right)$ \label{line:slackness}
        \State $x_i \gets x_{i+1} + \Delta_i$ \label{line:fwd_count}
    \EndFor
    \State \Return $\{x_0,\ldots, x_{S-1}\}$
\EndFunction
\end{algorithmic}
\end{algorithm}

\PHM{Dynamic adaption.} 
Upon detecting a communication delay exceeding the tolerance threshold
(given by \autoref{eq:absorb}), the system reconfigures the pipeline to
increase the slackness between the affected stages, aiming to
``absorb'' as many bubbles as possible. At this point, \emph{memory constraints can be
relaxed} as \SystemName{} offloads activations to host memory---which provides
significantly larger space than the device memory---and uses CPU-side RDMA for
data transfer to eliminate HOL blocking stalls (details in \S\ref
{sec:system_design}). Therefore, the optimal strategy is to maximize $\Delta_i$,
under virtually no memory limit.

\autoref{alg:adapt_fwds} implements this strategy. Starting backward from the
 last stage requiring only one warm-up forward (line~\ref
 {line:last_stage}), it recursively computes the desired warm-up count $x_i$ in
 the preceding stage (for-loop). Specifically, it uses profiled per-stage compute
 times and communication delays to determine the minimum required slackness
 $\Delta_i$ for all $i$ using \autoref{eq:absorb}, ensuring that $\Delta_i$
 is large enough to absorb the observed delay while clipping it by $N-2S$ to
 preserve enough forwards for other stages' warm-up (line~\ref
 {line:slackness}). Once $\Delta_i$ is computed, $x_i$ easily follows
 (line~\ref{line:fwd_count}).



\subsection{Full-Pipeline Orchestration}
\label{sec:pipeadapt}

With the warm-up forward count determined in each stage, we now construct the
complete pipeline execution schedule. Optimally orchestrating all operator
executions across stages and microbatches formulates a mixed-integer linear
programming (MILP) problem~\cite{guo2025deepseek,qi2023zero} and is
NP-hard~\cite{ullman1975np}. We hence turn to an efficient heuristic that
sequentially generates a pipeline schedule following its execution timeline,
discretized into multiple time steps. At each time step, the algorithm
does two things: (1) simulating pipeline execution and (2) making new scheduling
decisions.

\PHM{Simulation.}
First, it keeps track of the execution of previously scheduled operators
($F/B/W$) and updates their states in each stage---similar to running a
discrete-time simulation. It also maintains a list of \emph
{schedulable operators} for each stage and updates it accordingly (e.g., an
$F$ becomes schedulable in stage $S_i$ after its upstream $F$ completes in
$S_{i-1}$). 

\PHM{Operator selection.}
Second, for each idle stage, the algorithm makes new scheduling decisions
based on the updated system state. It chooses an operator from the
schedulable list following a \emph{two-phase operator selection policy}.
During the warm-up phase, each stage $S_i$ executes only forward operators
until reaching the assigned quota $x_i$ (computed by Algorithm~\ref
{alg:warmup_fwds} or~\ref{alg:adapt_fwds}), ensuring the desired straggler
resilience. After all warm-up forwards complete, the stage transitions into a
steady phase, in which it selects operators from the schedulable list in a
priority order of $B > F > W$. Specifically, backward input operators ($B$) are
prioritized to immediately free activation memory and propagate dependencies
upstream; forwards ($F$) are selected next, as they generate single
downstream dependencies; backward weight ($W$) operators have the lowest
priority, since they do not generate further dependencies and can be
scheduled opportunistically.

\PHM{Optimality and complexity.}
We will show in \S\ref{subsec:overhead} that this simple heuristic generates
pipeline schedules that closely approximate the optimum--obtained by solving
an MILP problem--when the simulation is configured with a fine-grained step
size $\delta$. In terms of the complexity, let $t_o$ be the longest operator execution
time, i.e., $t_o = \max_i(t^F_i, t^B_i, t^W_io)$. The algorithm completes in at most
$3NS\lceil t_o/\delta \rceil$ steps, as each operator (of which
there are $3N$ per stage over $S$ stages) is scheduled at most once per
$\delta$ interval. Each step involves $S$ constant-time policy evaluations,
resulting in an overall complexity of $O(NS^2\lceil t_o/\delta \rceil)$.

\section{Decoupled Data Plane via Comm.~Delegation}
\label{sec:system_design}

The effectiveness of the aforementioned pipeline adaptation algorithms (\S\ref
{sec:algorithm_design})---our first key design---is contingent on \emph
{eliminating head-of-line (HOL) blocking stalls}. In this section, we start
with a straw man solution and illustrate its ineffectiveness (\S\ref
{sec:straw_man_solution}). We then present our second key design to eliminate
HOL blocking stalls (\S\ref{sec:cpudelegation}), which additionally provides
fault tolerance to RNIC failures (\S\ref{sec:rnic_failure}).


\subsection{Straw Man Solution}
\label{sec:straw_man_solution}

Recall in \S\ref{sec:interference} that HOL blocking is caused by
sequential launching of communication and subsequent computation 
operations (\autoref{fig:profile_zb}). Therefore, the key to
avoiding HOL blocking is to \emph{decouple slow communication operations from
the compute sequence}, thereby ensuring that all compute kernels can be
launched without a delay. 

\begin{figure}[tb]
    \centering
    \includegraphics[width=0.95\linewidth]{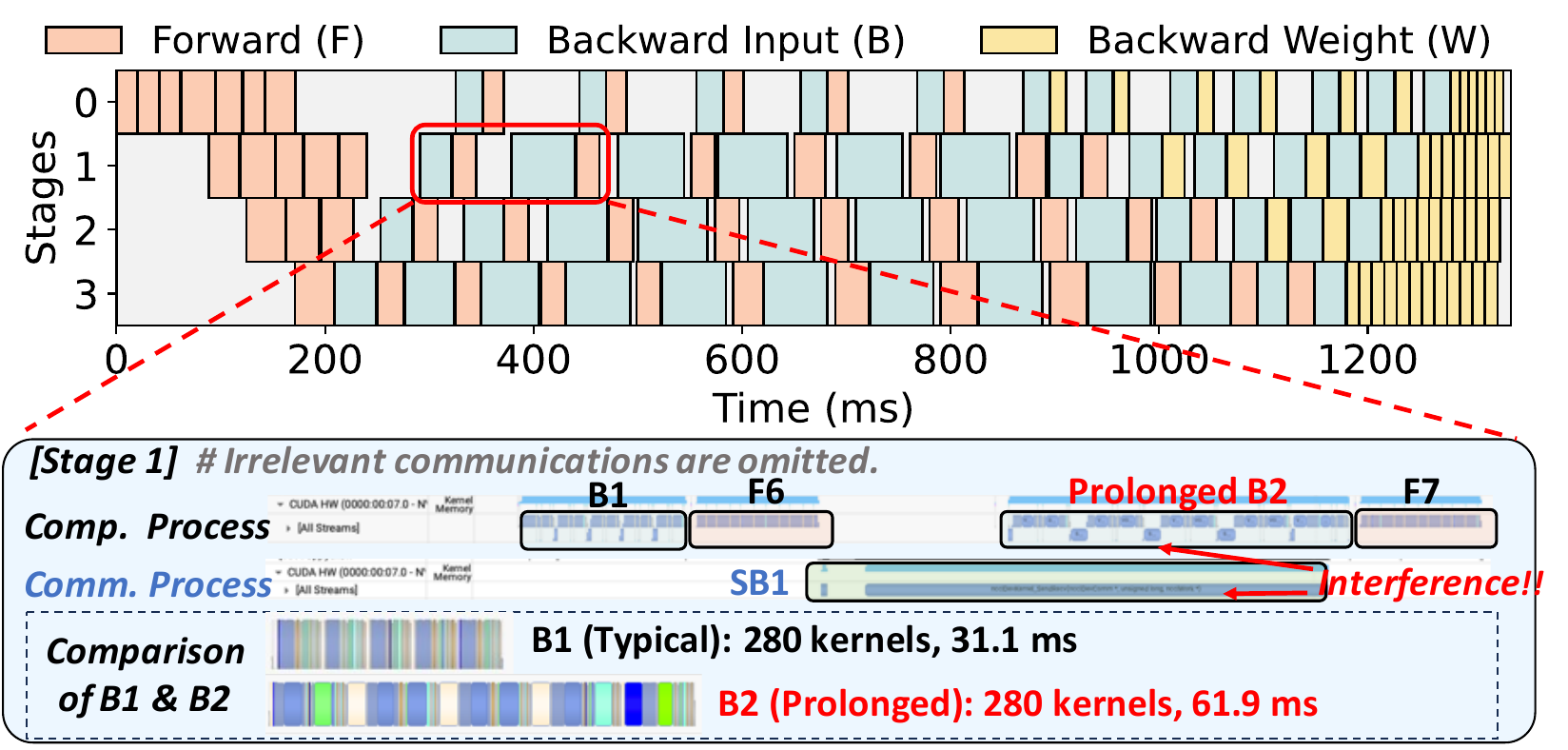}
    \caption{Naively adopting NCCL-based opportunistic communication~\cite{sourouri2014effective, athlur2022varuna} solves the blocking issue, but introduces severe interference to computation.}
    \label{fig:profile_gpu_delegate}
\end{figure}

A straw man solution is \emph{opportunistic communication}~\cite
{sourouri2014effective, athlur2022varuna}. It delegates communication
operations to some dedicated processes, allowing the main training process to
concentrate on computation. These dedicated communication processes
asynchronously retrieve data from shared buffers and transmit it to adjacent
pipeline stages via GPU-direct RDMA using NCCL. However, this approach
introduces significant interference to computation. As illustrated
in \autoref{fig:profile_gpu_delegate}, overlapping computation and
communication results in substantial kernel execution slowdowns. For instance,
stage-1's backward operation ($B_2$) increases from 31 ms to 61.9 ms when
overlapped with $SB_1$. Our profiling reveals that, although this approach
closes the kernel launch gaps, the runtime of individual kernels is greatly
prolonged: a GEMM kernel that typically finishes in 110~$\mu$s is stretched
to 2~ms under interference. As a result, despite being blocking-free, the
pipeline's end-to-end performance ends up no better than sequential
execution.



\subsection{CPU-based Communication Delegation}
\label{sec:cpudelegation}

\PHB{Key idea.}
While the NCCL-based straw man introduces severe interference, 
its delegation paradigm remains valid---provided we can
avoid such interference. This inspires us to offload activation and gradient
transfers from the GPU to \emph{host memory} upon straggler detection, and
perform send/receive operations using dedicated CPU-side delegate processes
to avoid interfering GPU computation. This design enables three key
benefits. \emph{First}, it fully decouples communication operations from GPU
kernel scheduling, preventing slow communication from stalling GPU
computation operations (blocking-free). \emph{Second}, offloading activation
and gradients to the host lifts the GPU memory pressure, enabling
orchestrating a memory-intensive pipeline schedule with more warm-up forwards
and larger slackness for enhanced straggler resilience (i.e., virtually no memory
constraint in Algorithm~\ref{alg:adapt_fwds}). \emph{Third}, it additionally provides RNIC fault
tolerance: in case of GPU-side RNIC failures, the host RNICs serves as a
backup.

\PHM{Design.}
In our design, the delegated communication path is activated only upon the detection of
communication delays. For each type of communication (e.g., send-forward),
the framework launches multiple CPU communication processes, each with its own
transmission queue. This multi-queue design ensures that the total data
consumption rate keeps pace with the data production rate of computation. For
example, if the GPU produces 8 activations per second while each
communication delegate can consume only 2 per second, at least 4 sending
queues are needed to avoid blocking and queue buildup.

\begin{figure}[tb]
    \centering
    \includegraphics[width=0.8\linewidth]{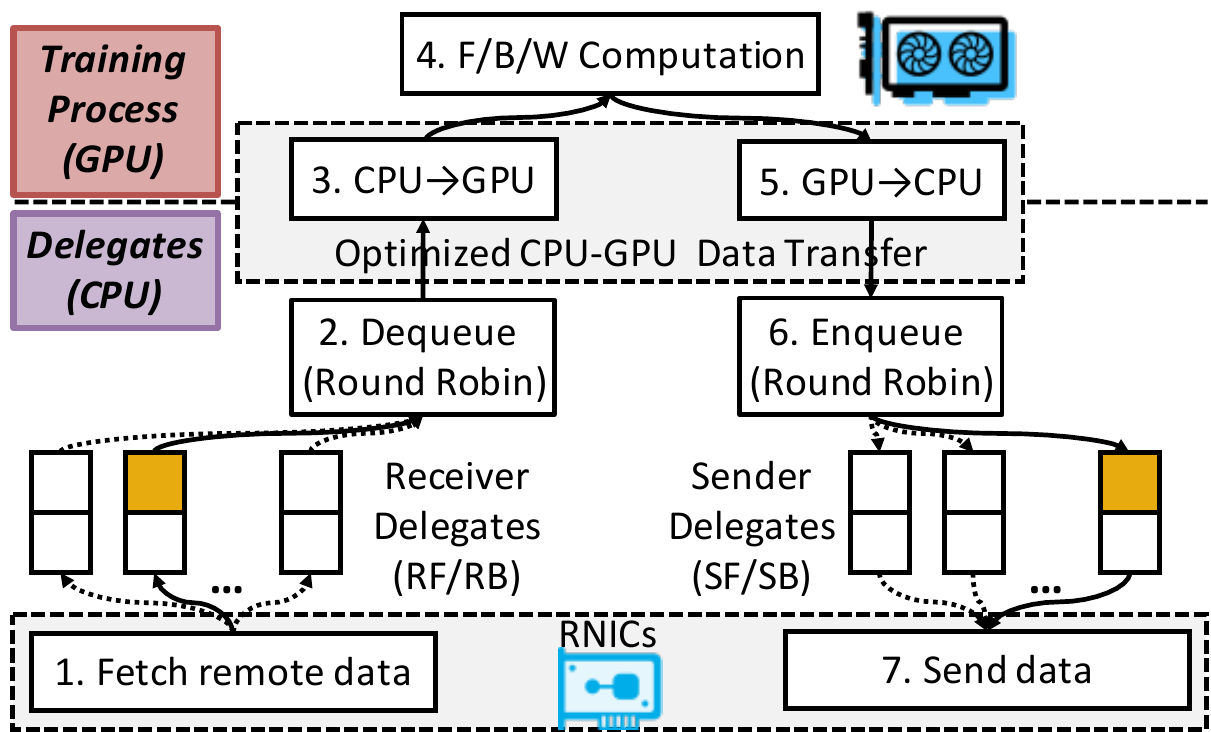}
    \caption{CPU-delegated data transmission path.}
    \label{fig:dele_path}
\end{figure}

\autoref{fig:dele_path} illustrates the data transmission path.
\textcircled{1} The receiver delegates eagerly fetch data from remote peers
 during training. The training process \textcircled{2} retrieves input data from
 a receiver queue in a \emph{round-robin} manner and \textcircled{3} copies it
 to GPU. The GPU \textcircled{4} computes the results, which are \textcircled
 {5} copied back to host and \textcircled{6} enqueued into the corresponding
 sender queue. \textcircled{7} Finally, the sender delegates send the results via
 CPU-side RDMA. 




\PHM{Optimizing Data transfer.}
As our design bypasses GPU-direct RDMA (GDR) and utilizes a slower CPU-side RDMA,
reducing the overhead of data movement between the host and the GPU becomes critical. To
minimize this overhead, we design a fine-grained data pipeline with optimized
CUDA kernels which move data asynchronously and only report to the
training process when the data is ready.

\begin{figure}[tb]
    \centering
    \includegraphics[width=\linewidth]{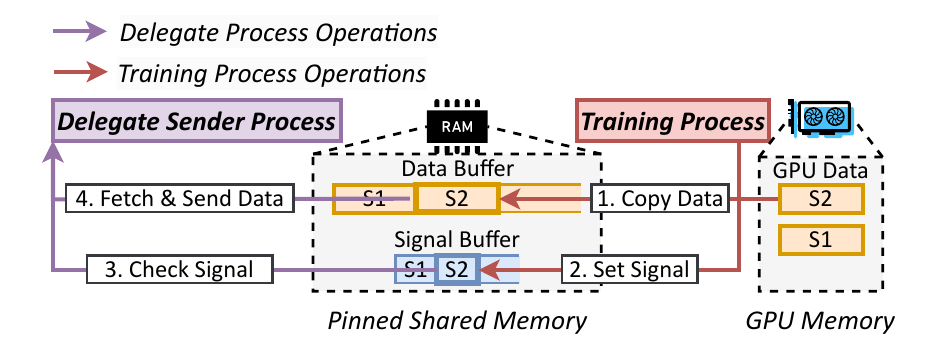}
    \caption{Optimized data transfer for sending data.}
    \label{fig:opt_send}
\end{figure}

Specifically, our optimization creates a pinned shared memory buffer for each
delegate, which allows faster GPU data access than pageable memory through
DMA and zero additional data copy between the delegate and the training
processes during IPC. For each replica of a communication type
(e.g., send-forward), both its training process and itself can
access a piece of pinned shared memory. When sending forward or backward, as
shown in \autoref{fig:opt_send}, the training process initiates two
sequential \texttt{cudaMemcpy()} operations in the same CUDA stream. It
first \textcircled{1} copies data from GPU to host and then \textcircled
{2} sets a copy completion signal to guarantee data integrity. After checking
the signal, the delegate process \textcircled{3} ensures copy completion
and \textcircled{4} sends data via RDMA.

\begin{figure}[tb]
    \centering
    \includegraphics[width=\linewidth]{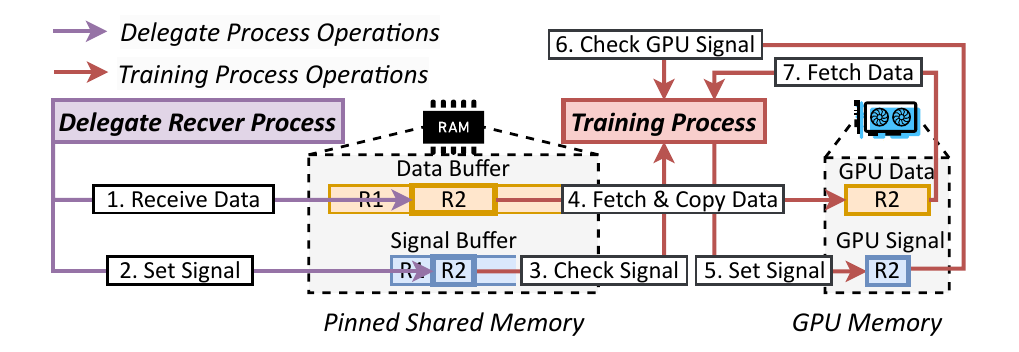}
    \caption{Optimized data transfer for receiving data.}
    \label{fig:opt_recv}
\end{figure}

When receiving forward or backward, as shown in \autoref
{fig:opt_recv}, once \textcircled{1} data is received from remote via RDMA,
\textcircled{2} the delegate process sets a signal in shared memory
indicating data ready. \textcircled{3} Meanwhile, the training process checks
this signal through busy waiting. After confirming, two sequential \texttt
{cudaMemcpy()} operations are initiated to first \textcircled{4} copy data
from host to GPU, followed by \textcircled{5} setting a signal in GPU memory
to acknowledge data copy completion. Once the training process \textcircled
{6} sees this data ready signal, the subsequent compute operators
can \textcircled{7} consume this data safely.

\subsection{Handling RNIC Failures}
\label{sec:rnic_failure}

During NCCL initialization, each GPU is assigned a dedicated RNIC to avoid
bandwidth contention within a node. Consequently, a single RNIC failure
during NCCL communication results in a connection loss or a timeout error,
even if the other RNICs are still well functioning. The GDR path is hence
vulnerable to RNIC failures. 

The CPU-based communication delegation provides an inherent tolerance to RNIC
failures as it bypasses the faulty GDR path. During training, each delegate
process can flexibly choose an RNIC for data transfer using Gloo~\cite
{gloo}. In the event of an RNIC failure, the delegate process reroutes the
affected communication traffic from the faulty RNIC to a healthy one to
continue data transfer, enabling uninterrupted training as opposed to
conventional checkpoint-and-restart failover.


\section{\SystemName{} Design and Implementation}
\label{sec:implmentation}

\begin{figure}[t]
    \centering
    \includegraphics[width=0.9\linewidth]{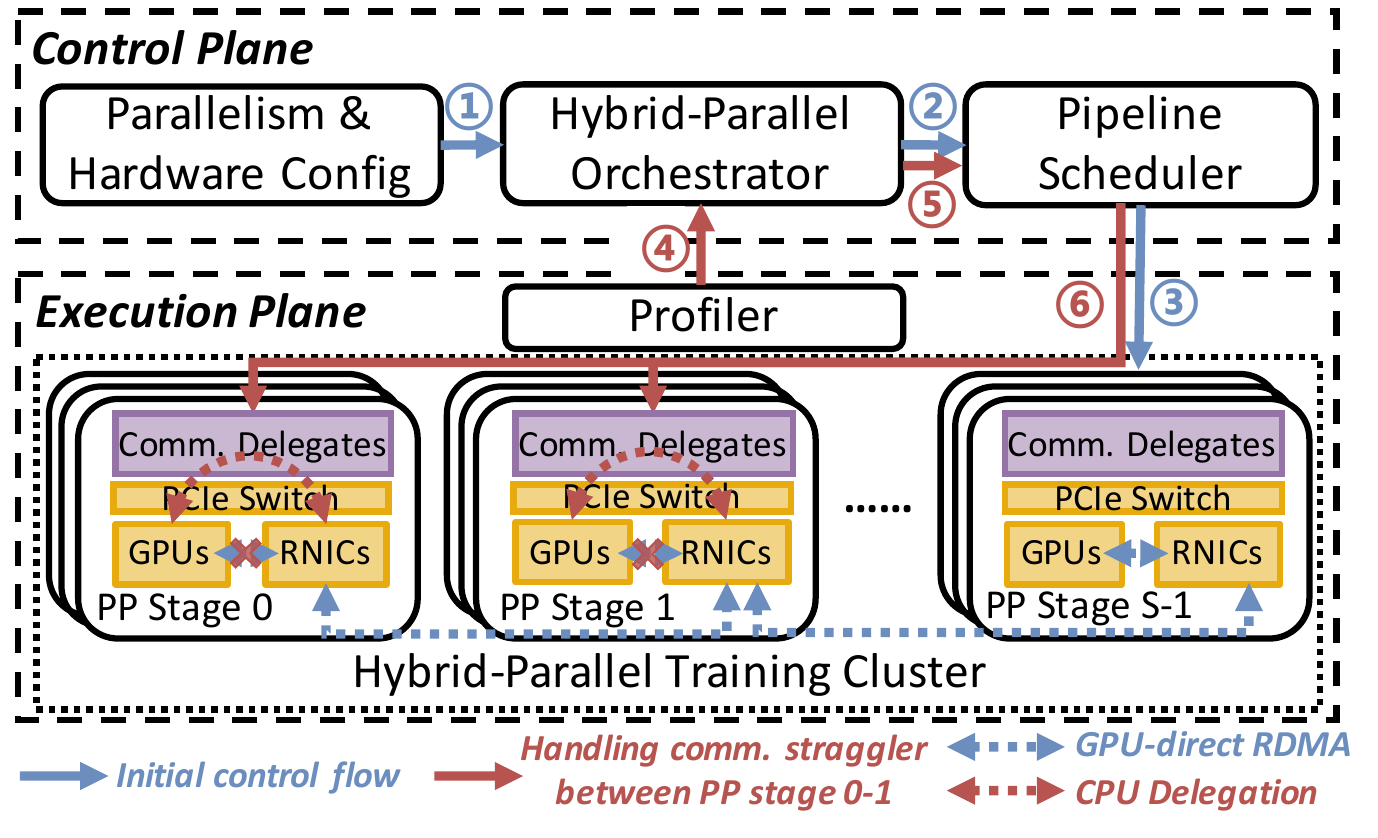}
    \caption{\SystemName system design.}
    \label{fig:sys_overview}
\end{figure}

\SystemName{} is an efficient parallel training system that integrates the
 two key designs described in \S\ref{sec:algorithm_design} and \S\ref
 {sec:system_design} to deliver robust resilience against communication
 stragglers. As illustrated in \autoref{fig:sys_overview}, it comprises 
 three main components: (1) a \textit{profiler} that continuously monitors each
 node's compute and communication performance, (2) a \textit
 {hybrid-parallel orchestrator} that dynamically determines the communication
 topology and constructs TP, DP, and PP communication groups based on runtime
 performance, and (3) a \textit{pipeline scheduler} that adaptively
 configures a resilient pipeline schedule against dynamic stragglers using
 the algorithms developed in \S\ref{sec:algorithm_design}.

\PHM{Initialization.} During system initialization, \textcircled{1} the orchestrator
 establishes the initial TP/DP/PP communication groups according to the
 parallelism strategy and hardware configuration (e.g., network
 topology). \textcircled{2} The scheduler then generates an initial pipeline
 schedule that provides uniform resilience to potential stragglers across all
 stages (\S\ref{sec:algorithm_design}). \textcircled{3} This schedule is
 deployed on the cluster, and the training starts with all inter-node
 communication performed via GPU-direct RDMA.

\PHM{Straggler mitigation.} Throughout training, the profiler continuously 
tracks communication and computation performance. Upon detecting slow
 communication (using detection techniques in~\cite
 {wu2024falcon,cui2025xputimer}), it reports this straggler event to the
 orchestrator (\textcircled{4}). If the affected link is between PP stages
 (e.g., stages 0 and 1), the orchestrator notifies the pipeline scheduler 
 (\textcircled{5}), which then adapts the pipeline schedule as described in 
 \S\ref{sec:algorithm_design}. The updated schedule is then
 deployed on the cluster, and CPU communication delegates are activated at the affected nodes to eliminate HOL blocking stalls (\textcircled{6}). If the slow link is
 part of a DP communication group, the orchestrator reconfigures the training
 topology to reassign this link for PP communication, effectively converting a DP
 straggler into a less detrimental PP straggler~\cite{wu2024falcon}, which is then
 addressed using the above mechanisms.

\PHM{Implementation.} \SystemName is implemented on top of Megatron-LM~\cite{narayanan2021megatron} and ZeroBubble~\cite{qi2023zero}, comprising 5.3K lines of code (LoC), primarily in Python,
with performance-critical data transfer kernels written in CUDA. The straggler detector and orchestrator are adapted from \GreyHound~\cite{wu2024falcon}, while the profiler leverages
CUDA Events and exposes performance profiles to the scheduler via Redis~\cite{redis}. For CPU-side communication, \SystemName{} utilizes Gloo~\cite{gloo} to facilitate RDMA data transfers.


\section{Evaluation}
\label{sec:eval}

In this section, we evaluate \SystemName to answer the following questions: (1) Does \SystemName effectively address dependency bubbles and HOL blocking stalls caused by PP stragglers (\S\ref{subsec:single})? (3) Does \SystemName also effectively handle DP stragglers (\S\ref{subsec:dpvspp})? (2) Can \SystemName generate an optimal schedule and delegate communication with acceptable overhead (\S\ref{subsec:overhead})? (4) How does \SystemName perform in large-scale pretraining in the presence of frequent communication stragglers and RNIC failures (\S\ref{subsec:largescale})?

\subsection{Experimental Setup}

\PHM{Cluster setup.} Our evaluation is conducted on a 128-GPU cluster, where each node is equipped with 8 NVIDIA H800 GPUs and 400 Gbps InfiniBand inter-node connections. We use CUDA 12.1 and NCCL 2.18.1 in the test environment.

\PHM{Baselines.} We evaluate \SystemName{} against four baselines.
\begin{enumerate}[noitemsep,nolistsep,,topsep=0pt,parsep=0pt,partopsep=0pt]
 \item 1F1B~\cite{narayanan2019pipedream} is a classic pipeline schedule with low bubble rate and controllable memory footprint.
 \item ZeroBubble (ZB)~\cite{qi2023zero} is a SOTA pipeline schedule that eliminates bubbles via decoupled backward passes.
 \item \GreyHound~\cite{wu2024falcon} migrates slow links to PP groups if stragglers occur on DP groups, but does not mitigate the stragglers' residual impact on pipeline execution.
 \item \SystemName-CPU only enables delegated communication (\S\ref{sec:system_design}) without pipeline adaptation.
\end{enumerate}

\PHM{Models and Parallelism.} We evaluate \SystemName using GPT-2 models of varying sizes, ranging from 7B to 140B parameters, on up to 128 GPUs across 16 nodes. The models and corresponding parallelism settings are given in \autoref{tab:model_parallelism_setting}.


\begin{table}[tb]
    \centering
    \footnotesize
    \begin{tabular}{c|c|c|c|c|c}
        \hline
        \textbf{Model Size} & 7B & 14B & 30B & 60B & 140B \\
        \hline
        \makecell[l]{\textbf{Parallelism}\\ \textbf{(TP, DP, PP)}} & (1, 1, 4) & (1, 1, 8) & (2, 4, 8) & (4, 2, 8) & (8, 2, 8)\\
        \hline
        \textbf{\#GPUs} & 4 & 8 & 64 & 64 & 128\\
        \hline
    \end{tabular}
    \caption{Models and corresponding 3D-parallelism settings.}
    \label{tab:model_parallelism_setting}
\end{table}

\subsection{Mitigating PP Stragglers}
\label{subsec:single}

\begin{figure}[tb]
    \centering
    \includegraphics[width=0.95\linewidth]{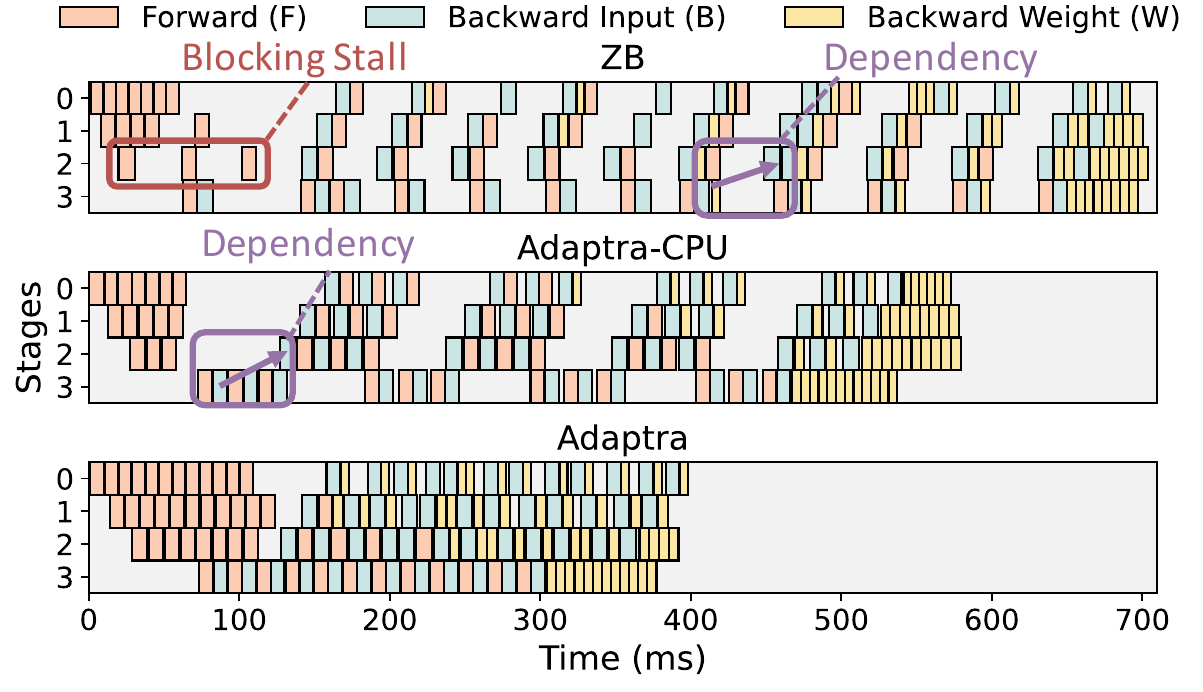}
    \caption{The actual execution of the schedule using \SystemName and other two baselines on a 7B model under 30 ms delay on the link between last two PP stages.}
    \label{fig:cmp}
\end{figure}

\PHB{Microbenchmark.} Before introducing end-to-end performance, we first demonstrate the behavior of \SystemName's two designs in addressing dependency bubbles and HOL blocking stalls using a GPT2-7B model. We inject a 30 ms delay between pipeline stages 2 and 3 and then profile the execution of each operator within an iteration using \texttt{cudaEvent}, with the execution timeline shown in \autoref{fig:cmp}.

In the original ZB execution, we observe significant bubbles between warm-up forward operators in stage 2 caused by HOL blocking stalls and dependency-bubbles exacerbate the performance. This compound effect extends the iteration time to 703 ms with 57.4\% bubble rate (1.9$\times$ longer than the normal execution). \SystemName-CPU retains the same pipeline schedule as ZB but activates CPU-based delegation, eliminating HOL blocking stalls by redirecting the original GPU-direct RDMA to CPU-based RDMA operations. As a result, the iteration time improves to 579 ms, with the bubble rate reduced to 48.8\% solely due to dependency issues. Further enabling pipeline adaptation (i.e., the complete \SystemName) reduces the iteration time to 398~ms (only 30~ms slowdown) and the bubble rate to only 25.3\%. This 1.76$\times$ improvement is achieved by increasing the slackness between stages 2 and 3 (i.e., $\Delta_2$), substantially enhancing the pipeline's resilience to the injected communication delays.

\begin{figure}[tb]
    \centering
    \includegraphics[width=0.95\linewidth]{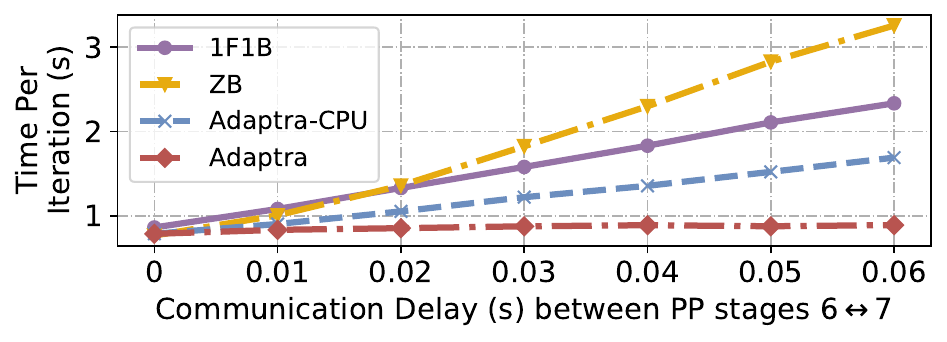}
    \caption{Sensitivity analysis of \SystemName and baselines using a 14B model under various delay values.}
    \label{fig:sensitivity}
\end{figure}

\PHM{Sensitivity analysis.} 
We assess the impact of delay values on end-to-end iteration times by
gradually increasing the latency between the last two stages of a 14B model
from 0 to 60 ms. As shown in \autoref{fig:sensitivity}, a 60~ms communication
delay slows down 1F1B, ZB, and \SystemName-CPU significantly by $2.70\times$,
$4.24\times$, and $2.15\times$, respectively. Notably, ZB, though achieving
better performance without stragglers, is more vulnerable to communication
delays due to its tightly coupled schedule. In contrast, \SystemName
consistently outperforms baseline systems with slightly increased iteration
times. Specifically, under a 60~ms delay, \SystemName measures a modest
slowdown of $1.13\times$ thanks to the two designs described in 
\S\ref{sec:algorithm_design} and \S\ref{sec:system_design}. Compared to ZB, 
switching to CPU-based communication delegation (\SystemName-CPU) mitigates 
the straggler impact by 48.1\%; this
impact is further alleviated by 24.6\% using pipeline adaptation. 

\begin{figure*}
    \centering
    \includegraphics[width=0.93\linewidth]{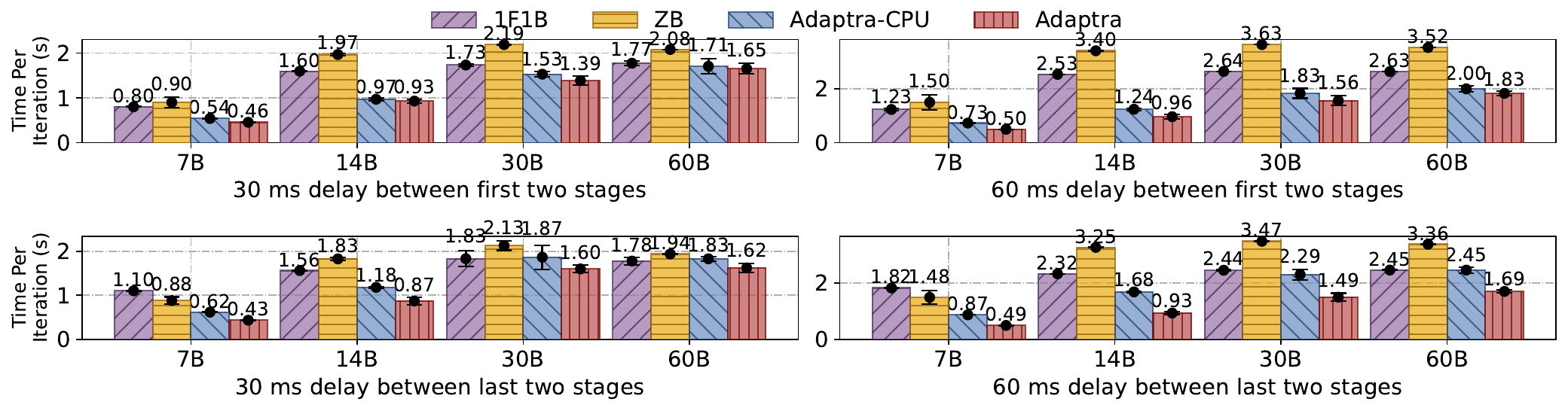}
    \caption{Evaluation on single inter-PP communication degradation under various model settings and delay locations.}
    \vspace{-0.5cm}
    \label{fig:single_delay}
\end{figure*}

\PHM{Resilience to single-link degradation.} 
We next evaluate \SystemName's performance in the presence of a single-link
straggler under various model and parallelism settings. As illustrated
in \autoref{fig:single_delay}, we inject a delay of 30~ms (or 60~ms) into a single communication
link between the first (or last) two PP stages. Across all model sizes from 7B
to 60B, \SystemName consistently outperforms baseline methods. In particular,
it achieves up to $3.71\times$ speedup over 1F1B (in the scenario of 
7B, 60~ms, last two stages) and $3.49\times$ speedup over ZB (14B, 60~ms, last). When the
communication delay increases from 30~ms to 60~ms, the average iteration time measured across
all models increases by only $1.07\times$ using \SystemName, compared
to $1.49\times$, $1.69\times$, and $1.30\times$ using 1F1B, ZB,
and \SystemName-CPU, respectively. These results suggest that
dependency bubbles account for 39\% of slowdown, while HOL blocking stalls
contribute 23\% on average. \SystemName effectively mitigates both issues.

Note that the straggler location also matters, as delays between the last two
PP stages are difficult to hide in the original ZB schedule due to their tight
dependencies. \SystemName effectively addresses this issue through
pipeline adaptation: it improves the average iteration time by 19.3\%
compared to \SystemName-CPU under a 60~ms delay between the first two stages;
this gain increases to 38.4\% when the same delay occurs between the last two stages.

\begin{figure}[tb]
    \centering
    \includegraphics[width=0.95\linewidth]{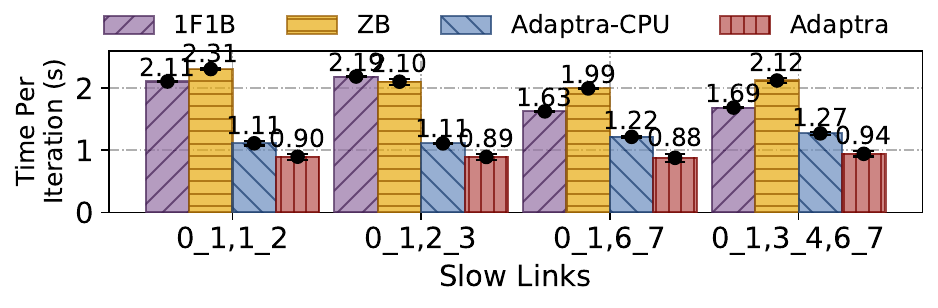}
    \caption{Iteration times of a 14B model under multiple simultaneous inter-PP communication stragglers.}
    \label{fig:multi_delay}
\end{figure}

\PHM{Multi-link degradation.}
To evaluate \SystemName under more complex straggler conditions, we configure multiple
stragglers in the 14B setting with 30~ms delays occurring on two adjacent links 
(0-1 and 1-2), two skip links (0-1 and 2-3), first-and-last links (0-1 and 6-7), 
and three skip links (0-1, 3-4, 6-7), respectively. As detailed in \autoref
{fig:multi_delay}, \SystemName reduces the iteration time by 51.6\% and
57.5\% on average across the four settings, compared to 1F1B and ZB. Importantly,
adaptive pipeline reconfiguration improves the average performance by 23.1\%
(comparing to \SystemName-CPU), further validating the effectiveness
of \SystemName's scheduling algorithm under complicated delay conditions.

\subsection{Mitigating DP Stragglers}
\label{subsec:dpvspp}

\begin{figure}[tb]
    \centering
    \includegraphics[width=0.95\linewidth]{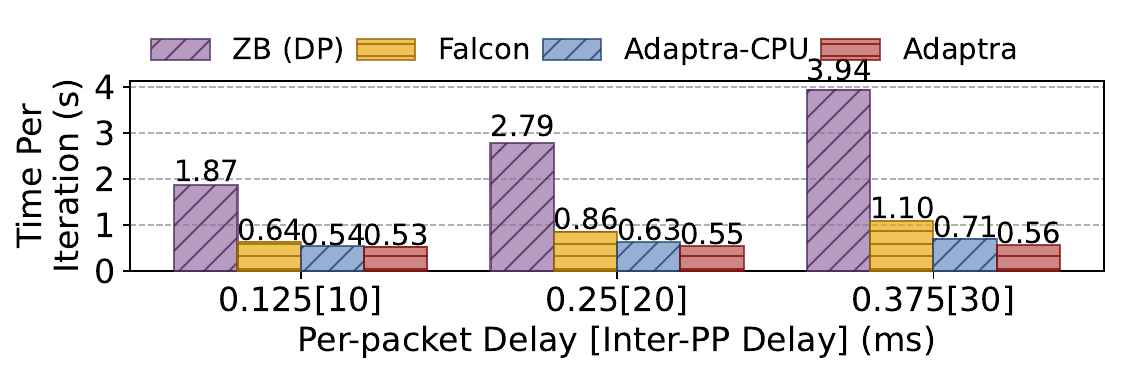}
    \caption{A communication straggler in DP group introduces significant baseline degradation. Both \GreyHound~\cite{wu2024falcon} and \SystemName migrates this link to PP groups, while \SystemName further optimizes the residual impacts.}
    \label{fig:slowdp}
\end{figure}

Communication stragglers can also occur between DP groups. We
evaluate \SystemName in this scenario using a GPT2-7B model on 8 nodes with
ZB scheduling under a configuration of (1 TP, 2 DP, 4 PP). We inject
per-packet delays of 0.125/0.25/0.375 ms (10/20/30 ms inter-PP stage latency)
into a designated link. As shown in \autoref{fig:slowdp}, when this link is
for DP communications, the baseline ZB degrades by $8.04\times$, far
exceeding the impact of PP communication stragglers. In response,
both \SystemName and \GreyHound~\cite{wu2024falcon} mitigate this by
reassigning the slow link to PP communication groups, leading to
$3.6\times$ improvement. \SystemName further mitigates the residual PP straggler through
pipeline adaptation and CPU-delegated communications, achieving $1.96\times$
additional speedup over \GreyHound.

\subsection{Optimality and Overhead}
\label{subsec:overhead}

\begin{figure}[tb]
    \centering
    \includegraphics[width=0.95\linewidth]{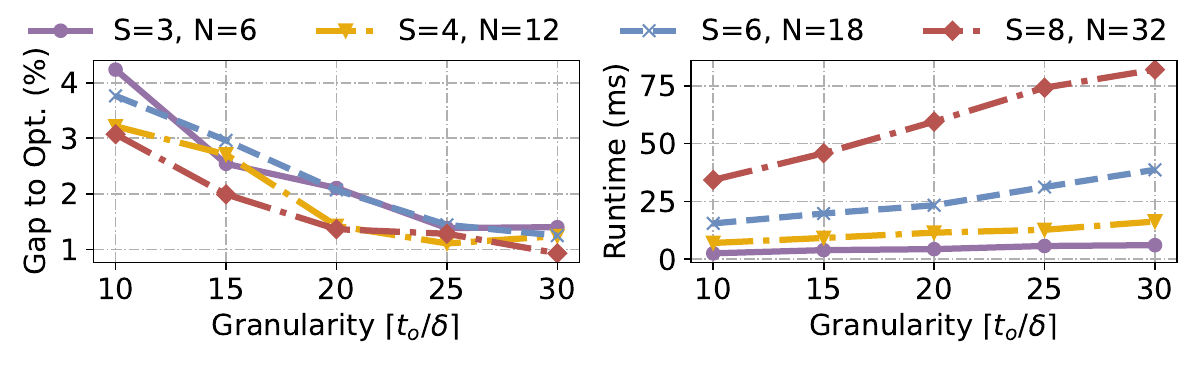}
    \caption{The relative error to optimal solution and solving time of \SystemName's scheduler, where it achieves an near-optimal solution within 1\% error in 0.1 seconds.}
    \label{fig:schedule_cost}
\end{figure}

\PHB{Performance and overhead of pipeline scheduling.} We evaluate the
 performance of our pipeline scheduling algorithm described in \S\ref
 {sec:algorithm_design} against the optimal schedule obtained by formulating
 an MILP problem. We consider four pipeline configurations (with 3-8 stages
 and 6-32 microbatches) with randomly generated profiles. As illustrated
 in \autoref{fig:schedule_cost}, configuring a smaller time step $\delta$
 (higher $\lceil t_o/\delta\rceil$) for fine-grained simulation narrows the
 gap between the generated schedule and the optimum, at a cost of increased
 schedule generation time. In all settings, the gap, measured by the
 relative difference in execution time, is less than 1\% when running
 at a fine granularity of $\lceil t_o/\delta\rceil=30$. These near-optimal schedules 
 are generated in less than 100~ms, as opposed to computing the optimal schedules
 that requires solving an MILP in hours.

\PHM{Overhead of delegation.}
We stress-test the delegated path across models from 7B to 60B parameters. Under a \emph{worst-case scenario} where all pipeline communications are forcibly routed through CPU delegates (All-CPU), \SystemName maintains comparable performance (<$5\%$ overhead) to the GPU-direct RDMA (GDR) baselines (ZB and 1F1B) for models with $\le30\text{B}$  parameters. Even at 60B scale, All-CPU introduces only 17\% additional iteration time compared to GDR.

\begin{figure}[tb]
    \centering
    \includegraphics[width=0.95\linewidth]{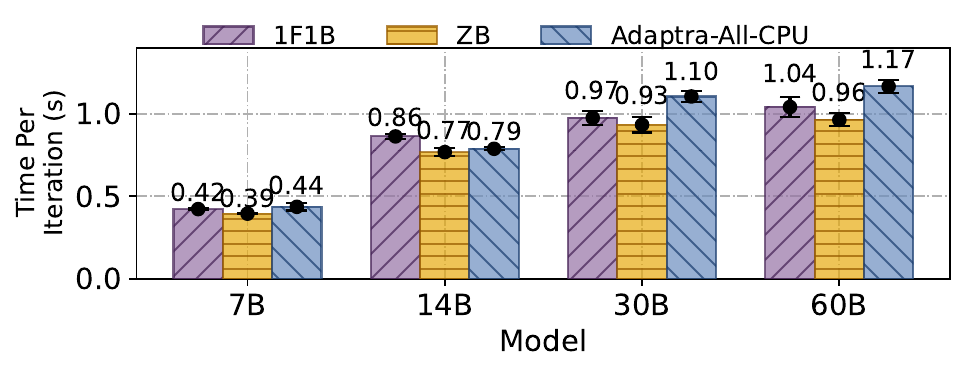}
    \caption{Worst-case overhead of \SystemName's delegated communication w/o presence of communication stragglers.}
    \label{fig:overhead_nodelay}
\end{figure}



\subsection{Large-Scale Evaluation}
\label{subsec:largescale}

We further evaluate \SystemName on a 140B model using 128 GPUs across 16
nodes. We construct a 1,200-iteration trace which includes 9 communication
straggler events in PP groups (with 1–3 concurrent slow links per event,
20–60~ms latency) and one RNIC failure (details given in Appendix). Each
straggler event starts at a certain iteration and lasts for 70 iterations
before reverting to normal operation. During the RNIC failure, 1F1B and
ZB perform checkpoint-and-restart, while \SystemName runs uninterruptedly via
delegation. 

\begin{figure}[tb]
    \centering
    \includegraphics[width=\linewidth]{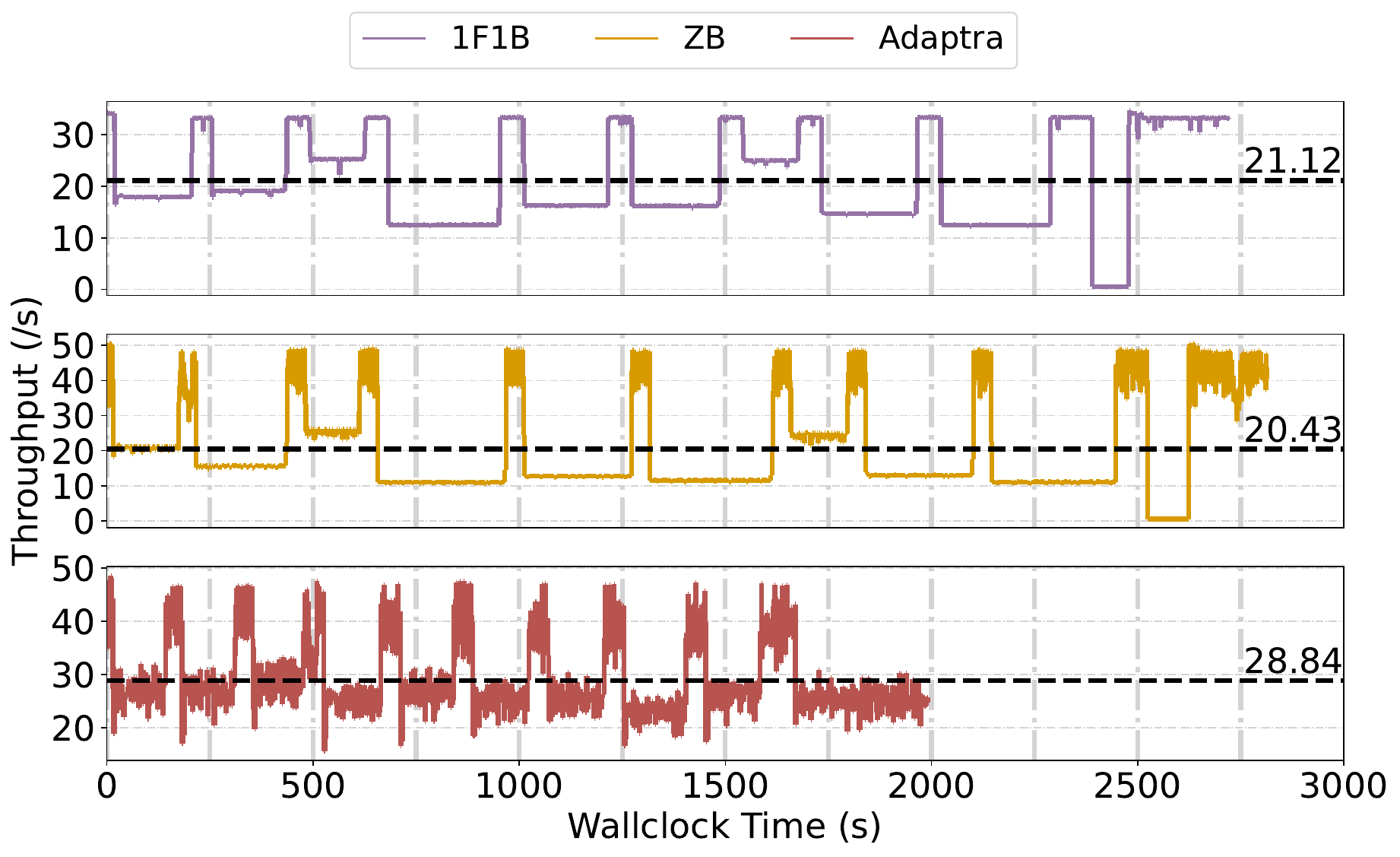}
    \caption{Throughput of \SystemName of pretraining a 140B model on 128 NVIDIA H800 GPUs against baselines, where \SystemName improves the throughput by up to $1.41\times$.}
    \label{fig:largescale}
\end{figure}

As shown in \autoref{fig:largescale}, without delays, 1F1B’s inherent pipeline bubbles caused lower throughput compared to ZB and \SystemName{}. However, when the latency value is relatively large, the
throughput of ZB decreases to around 10 samples per second, even worse than
that of 1F1B. In contrast, by resolving HOL blocking stalls and dependency
bubbles, \SystemName maintains more than 20 samples per second most of the
time, significantly surpassing the two baselines. Over the full training
cycle, \SystemName{} completed in 33.2 minutes, outperforming 1F1B
(45.4 minutes) and ZB (46.9 minutes) by $1.37\times$ and $1.41\times$,
respectively. This demonstrates \SystemName{}'s ability to maintain robust performance under simultaneous stragglers and hardware failures.

\section{Related Work}
\label{sec:related_work}

\PHB{Reliability issues in training.} Abundant studies address training crashes using checkpoints~\cite{mohan2021checkfreq, wang2023gemini}, redundant computations~\cite{thorpe2023bamboo}, and elastic frameworks~\cite{oobleck, thorpe2023bamboo, mohan2021checkfreq,wang2023gemini}. However, seldom existing works address communication stragglers, while they have been identified in several reports~\cite{dubey2024llama, jiang2024megascale},. Malleus~\cite{li2024malleus} solely mitigates compute stragglers without considering communications. Falcon~\cite{wu2024falcon} addresses slow communication by shifting the degraded links to PP groups without optimizing the residual impact. Crux~\cite{cao2024crux}'s communication-aware scheduling tries to reduce the occurrence probability of stragglers, yet not mitigating their impacts.

\PHM{Communication optimizations for training.} Communication optimizations span three critical layers. Infrastructure-level efforts like Alibaba HPN~\cite{qian2024alibaba} and Megascale~\cite{jiang2024megascale} propose specialized network topologies for training clusters. At the library level, TACCL~\cite{shah2023taccl}, ACCL~\cite{dong2021accl}, and MSCCL~\cite{cowan2023mscclang} develop optimized communication primitives for collective operations. Framework-level approaches including Megatron-LM~\cite{narayanan2021megatron}, Varuna~\cite{athlur2022varuna}, and DeepEP~\cite{deepep2025} enhance computation-communication overlap in hybrid-parallel settings.

\PHM{Pipeline parallelism optimizations.}  Pipeline scheduling remains challenging in distributed training. Classic approaches like Gpipe~\cite{huang2019gpipe} and 1F1B~\cite{narayanan2019pipedream} achieve comparable bubble rates, with 1F1B additionally reducing memory usage.  Recent advances address distinct dimensions: Interleaved 1F1B~\cite{narayanan2021megatron} reduces bubbles via introducing virtual stages, while ZeroBubble~\cite{qi2023zero} eliminates them through decoupled backward passes. For specialized architectures, DualPipe~\cite{guo2025deepseek} optimizes MoE training pipelines and RLHFuse~\cite{zhong2024rlhfuse} tailors for RLHF workloads. In terms of improving reliability, Recycle~\cite{gandhi2024recycle} employs precomputed schedules with microbatch rerouting for fail-stop recovery, whereas SDPipe~\cite{miao2023sdpipe} trades training accuracy for computation straggler-resilience.

\section{Conclusion}
\label{sec:conclusion}

This paper presents \SystemName{}, a system designed for efficient hybrid-parallel training
in the presence of communication stragglers. \SystemName employs dynamic pipeline adaption to minimize dependency bubbles and CPU-delegated communication to eliminate head-of-line blocking stalls. Experiments demonstrate that \SystemName achieves 1.2-3.5$\times$ speedup over SOTA baselines under communication stragglers, and $1.41\times$ higher throughput with zero restart overhead upon RNIC failures.



\clearpage
\bibliographystyle{plain}
\bibliography{sample}

\clearpage
\section*{Appendix}
\label{sec:appendix}
\subsection*{Proof of the Lemma in \S~\ref{sec:analysis}}
\label{apdx:prooflemma}
Assume for contradiction that $x_i < x_{i+1}$ for some stage $S_i$.
Since the warm-up phase of $S_i$ ends after $F_{i, x_i}$, $F_{i, x_i+1}$ must belong to the steady phase of $S_i$, preceded by $B_{i, 1}$, i.e., 
\begin{equation}
    \begin{aligned}
        F_{i, x_i} \prec B_{i, 1} \prec F_{i, x_i + 1}.
    \end{aligned}    
\end{equation}
However, $B_{i+1, 1}$ can only occur after $S_{i+1}$’s warm-up phase ends with $F_{i+1, x_{i+1}}$. Furthermore, due to pipeline dependencies, $F_{i+1, x_{i+1}}$ depends on $F_{i, x_{i+1}}$. Thus:
\begin{equation}
    \begin{aligned}
        F_{i, x_{i+1}} \prec F_{i+1, x_{i+1}} \prec B_{i+1, 1}.
    \end{aligned}    
\end{equation}
Combining these inequalities, we have:  
\begin{equation}
    \begin{aligned}
        B_{i, 1} \prec F_{i, x_{i+1}} \prec B_{i+1, 1}.
    \end{aligned}    
\end{equation}
This implies that $B_{i, 1}$ starts before $B_{i+1,1}$ which violates the data dependency of backward between $S_i$ and $S_{i+1}$. Hence, $x_i \geq x_{i+1}$ must hold for all $1 \leq i \le S$.  \unskip\nobreak\hfill $\square$

\subsection*{Scheduling Algorithm in \S~\ref{sec:pipeadapt}}
We present the detailed algorithm of generating the full pipeline schedule, which employs a discrete-time simulator and a localized operation selection policy described in \S~\ref{sec:pipeadapt}. \autoref{alg:deltai_policy} presents our two-stage operator selection policy, which guarantees the slackness on the communication stragglers and minimizes the bubble rate for subsequent steady and cool-down phases. \autoref{alg:schedule} then demonstrates the implementation of pipeline simulation, which asks the policy for operator selection and propagates scheduable operators to the upstream and downstream stages.

\begin{algorithm}[htbp]
\small
\caption{Operator Selection Policy}
\label{alg:deltai_policy}
\begin{algorithmic}[1]
\Require Current stage $i$ ($0\le i < S$), available operator sets $\mathcal{A}_i$ at current step of $S_i$, required warm-up forward counts $x_i$.
\Ensure An operator $o \in \{F,B,W\}$ or None if no operators should be executed.
\Function{SelectOp}{$i, \mathcal{A}_i, x_i$}
    \If {$\mathcal{A}_i = \emptyset$}
        \State \Return None
    \EndIf
    \State $\triangleright$ \textit{\textcolor{blue}{Warm-up phase, do $x_i$ forwards at first.}}
    \If{$x_i > 0$}
        \If {$F\in \mathcal{A}_i$}
            \State $x_i \gets x_i - 1$
            \State \Return $\mathcal{A}_i.\text{pop}(F)$
        \Else
            \State \Return None
        \EndIf
    \EndIf
    \State $\triangleright$ \textit{\textcolor{blue}{Execution priority after warm-up: $B>F>W$.}}
    \State $Pri \gets \{B:3, F:2, W:1\}$
    \State $\triangleright$ \textit{\textcolor{blue}{Find an available operator with the highest priority.}}
    \State \Return $\mathcal{A}_i.\text{pop}(\max_{o\in \mathcal{A}_i}(Pri[o.type]))$
\EndFunction
\end{algorithmic}
\end{algorithm}

\begin{algorithm}[htbp]
\small
\caption{Full Pipeline Schedule Generation}
\label{alg:schedule}
\begin{algorithmic}[1]
\Require Number of stages $S$, number of microbatches $N$, per-stage compute times $\{t^F_i\},\{t^B_i\},\{t^W_i\}$, inter-PP communication times $\{c_i\}$, warm-up forward counts $\{x_i\}$, time step $\delta$.
\Ensure A pipeline schedule $\mathcal{X}$.
\Function{Schedule}{$S, N, \{t^F_i\},\{t^B_i\},\{t^W_i\},\{x_i\}, \{c_i\}, \delta$}
    \State $\mathcal{X}\gets [\emptyset] \times S$ \Comment{\textit{\textcolor{blue}{Initialize an empty schedule.}}}
    \State $\mathcal{A} \gets [\emptyset]\times S$ \Comment{\textit{\textcolor{blue}{Current available operators.}}}
    \State $\mathcal{A}_0 \gets [F]\times N$ \Comment{\textit{\textcolor{blue}{Initially, $F$'s are available for $S_0$.}}}
    \State $t \gets 0$
    \While {$\exists\ A\in\mathcal{A}, A\ne \emptyset$}
        \For {$i \in [0\ldots S - 1]$}
            \If {not $i$.busy()}
                 \State $\triangleright$ \textit{\textcolor{blue}{Decide the next operator to execute.}}
                \State $o \gets \textsc{SelectOp}(i, \mathcal{A}_i, x_i)$
                \State $i.\text{execute}(o, duration={t}^{o.type}_i)$
                \State $\triangleright$ \textit{\textcolor{blue}{Add dependent operators after execution.}}
                \If{$o.type = F$ and $i \ne S-1$}
                    \State $\mathcal{A}_{i+1}.\text{append}(F(ready=t+c_i))$
                \ElsIf{$o.type = B$ and $s \ne 0$}
                    \State $\mathcal{A}_{i-1}.\text{append}(B(ready=t+c_{i-1}))$
                    \State $\mathcal{A}_{i-1}.\text{append}(W(ready=t+c_{i-1}))$
                \EndIf
                \State $\mathcal{X}_i\gets \mathcal{X}_i\cup \{o\}$ \Comment{\textit{\textcolor{blue}{Update schedule.}}}
            \EndIf
        \EndFor
        \State $t \gets t + \delta$
    \EndWhile
    \State \Return $\mathcal{X}$
\EndFunction
\end{algorithmic}
\end{algorithm}

\subsection*{Injection Trace in \S~\ref{subsec:largescale}}
We inject nine communication stragglers and one RNIC failure in the large-scale experiment (\S~\ref{subsec:largescale}). Their durations and severity (i.e., inter-stage communication latencies) are provided in \autoref{tab:trace}. The $\infty$ latency at step 1030 indicates a single RNIC failure, where a checkpoint-and-restart process is required to resume training for 1F1B or ZB schedule, while it can be smoothly handled by \SystemName's delegated path.

\begin{table}[htbp]
    \centering
    \footnotesize
    \begin{tabular}{c|ccccc}
        \textbf{EventID} & 0 & 1 & 2 & 3 & 4 \\
        \hline
        \textbf{Duration} & 15-85 & 120-190 & 230-300 & 340-410 & 450-520 \\
        \textbf{Stages} & 
        $2\leftrightarrow 3$ & 
        \makecell[l]{$0\leftrightarrow 1$\\ $5\leftrightarrow 6$} & 
        $6\leftrightarrow 7$ & 
        \makecell[l]{$2\leftrightarrow 3$\\ $3\leftrightarrow 4$\\ $6\leftrightarrow 7$} & 
        $5\leftrightarrow 6$\\
        \textbf{Lat. (ms)} & 30 & 40 & 20 & 50 & 60 \\
    \end{tabular}\\[1em]
    \begin{tabular}{c|ccccc}
        \textbf{EventID} & 5 & 6 & 7 & 8 & 9 \\
        \hline
        \textbf{Duration} & 560-630 & 670-740 & 780-850 & 890-960 & 1030 \\
        \textbf{Stages} & 
        \makecell[l]{$1\leftrightarrow 2$\\ $6\leftrightarrow 7$} & 
        \makecell[l]{$0\leftrightarrow 1$\\ $4\leftrightarrow 5$} & 
        \makecell[l]{$0\leftrightarrow 1$\\ $1\leftrightarrow 2$\\ $2\leftrightarrow 3$} & 
        \makecell[l]{$4\leftrightarrow 5$\\ $5\leftrightarrow 6$} & 
        2 \\
        \textbf{Lat. (ms)} & 60 & 20 & 40 & 50 & $\infty$ \\
    \end{tabular}
    \caption{Injected trace of communication stragglers and RNIC crashes used in \S~\ref{subsec:largescale}.}
    \label{tab:trace}
\end{table}


\end{document}